\documentclass{aastex631}

\received{\today}
\revised{\today}
\accepted{\today}

\submitjournal{ApJ}

\shorttitle{Analysis of 3HWC J1928+178, 3HWC J1930+188 and HAWC J1932+192}
\shortauthors{The HAWC collaboration}
\usepackage{textcomp}
\usepackage{txfonts}
\usepackage[utf8]{inputenc}
\usepackage{multirow}
\DeclareUnicodeCharacter{2212}{-}

\begin{document}

\title{Detailed Analysis of the TeV $\gamma$-Ray Sources 3HWC J1928+178, 3HWC J1930+188, and the New Source HAWC J1932+192}

\correspondingauthor{Armelle Jardin-Blicq}
\email{armelle.jardin-blicq@mpi-hd.mpg.de}
\correspondingauthor{Vincent Marandon}
\email{vincent.marandon@mpi-hd.mpg.de}

\author[0000-0003-0197-5646]{A.~Albert}
\affiliation{Physics Division, Los Alamos National Laboratory, Los Alamos, NM, USA }

\author{R.~Alfaro}
\affiliation{Instituto de F\'{i}sica, Universidad Nacional Autónoma de México, Ciudad de Mexico, Mexico }

\author{C.~Alvarez}
\affiliation{Universidad Autónoma de Chiapas, Tuxtla Gutiérrez, Chiapas, Mexico}

\author{J.C.~Arteaga-Velázquez}
\affiliation{Universidad Michoacana de San Nicolás de Hidalgo, Morelia, Mexico }

%\author[0000-0002-3032-663X]{K.P.~Arunbabu}
%\affiliation{Instituto de Geof\'{i}sica, Universidad Nacional Autónoma de México, Ciudad de Mexico, Mexico }

\author[0000-0002-4020-4142]{D.~Avila Rojas}
\affiliation{Instituto de F\'{i}sica, Universidad Nacional Autónoma de México, Ciudad de Mexico, Mexico }

\author[0000-0002-2084-5049]{H.A.~Ayala Solares}
\affiliation{Department of Physics, Pennsylvania State University, University Park, PA, USA }

\author[0000-0002-5529-6780]{R.~Babu}
\affiliation{Department of Physics, Michigan Technological University, Houghton, MI, USA }

%\author[0000-0003-0477-1614]{V.~Baghmanyan}
%\affiliation{Institute of Nuclear Physics Polish Academy of Sciences, PL-31342 IFJ-PAN, Krakow, Poland }

\author[0000-0003-3207-105X]{E.~Belmont-Moreno}
\affiliation{Instituto de F\'{i}sica, Universidad Nacional Autónoma de México, Ciudad de Mexico, Mexico }

%\author[0000-0001-5537-4710]{S.Y.~BenZvi}
%\affiliation{Department of Physics \& Astronomy, University of Rochester, Rochester, NY , USA }

\author[0000-0002-5493-6344]{C.~Brisbois}
\affiliation{Department of Physics, University of Maryland, College Park, MD, USA }

\author[0000-0002-4042-3855]{K.S.~Caballero-Mora}
\affiliation{Universidad Autónoma de Chiapas, Tuxtla Gutiérrez, Chiapas, Mexico}

\author[0000-0003-2158-2292]{T.~Capistrán}
\affiliation{Instituto de Astronom\'{i}a, Universidad Nacional Autónoma de México, Ciudad de Mexico, Mexico }

\author[0000-0002-8553-3302]{A.~Carramiñana}
\affiliation{Instituto Nacional de Astrof\'{i}sica, Óptica y Electrónica, Puebla, Mexico }

\author[0000-0002-6144-9122]{S.~Casanova}
\affiliation{Institute of Nuclear Physics Polish Academy of Sciences, PL-31342 IFJ-PAN, Krakow, Poland }

\author{O.~Chaparro-Amaro}
\affiliation{Centro de Investigaci\'on en Computaci\'on, Instituto Polit\'ecnico Nacional, M\'exico City, Mexico.}

\author[0000-0002-7607-9582]{U.~Cotti}
\affiliation{Universidad Michoacana de San Nicolás de Hidalgo, Morelia, Mexico }

\author[0000-0002-1132-871X]{J.~Cotzomi}
\affiliation{Facultad de Ciencias F\'{i}sico Matemáticas, Benemérita Universidad Autónoma de Puebla, Puebla, Mexico}

\author[0000-0002-7747-754X]{S.~Coutiño de León}
\affiliation{Department of Physics, University of Wisconsin-Madison, Madison, WI, USA }

\author[0000-0001-9643-4134]{E.~De la Fuente}
\affiliation{Departamento de F\'{i}sica, Centro Universitario de Ciencias Exactase Ingenierias, Universidad de Guadalajara, Guadalajara, Mexico }
\affiliation{Institute for Cosmic Ray Research, University of Tokyo, 277-8582 Chiba, Kashiwa, Kashiwanoha, 5 Chome-1-5, Japan}

\author[0000-0002-8528-9573]{C.~de León}
\affiliation{Universidad Michoacana de San Nicolás de Hidalgo, Morelia, Mexico }

\author[0000-0001-8487-0836]{R.~Diaz Hernandez}
\affiliation{Instituto Nacional de Astrof\'{i}sica, Óptica y Electrónica, Puebla, Mexico }

\author[0000-0002-0087-0693]{J.C.~Díaz-Vélez}
\affiliation{Departamento de F\'{i}sica, Centro Universitario de Ciencias Exactase Ingenierias, Universidad de Guadalajara, Guadalajara, Mexico }

\author[0000-0001-8451-7450]{B.L.~Dingus}
\affiliation{Physics Division, Los Alamos National Laboratory, Los Alamos, NM, USA }

\author[0000-0002-2987-9691]{M.A.~DuVernois}
\affiliation{Department of Physics, University of Wisconsin-Madison, Madison, WI, USA }

\author[0000-0003-2169-0306]{M.~Durocher}
\affiliation{Physics Division, Los Alamos National Laboratory, Los Alamos, NM, USA }

%\author[0000-0003-2338-0344]{R.W.~Ellsworth}
%\affiliation{Department of Physics, University of Maryland, College Park, MD, USA }

\author[0000-0001-5737-1820]{K.~Engel}
\affiliation{Department of Physics, University of Maryland, College Park, MD, USA }

\author[0000-0001-7074-1726]{C.~Espinoza}
\affiliation{Instituto de F\'{i}sica, Universidad Nacional Autónoma de México, Ciudad de Mexico, Mexico }

\author[0000-0002-8246-4751]{K.L.~Fan}
\affiliation{Department of Physics, University of Maryland, College Park, MD, USA }

\author{M.~Fernández Alonso}
\affiliation{Department of Physics, Pennsylvania State University, University Park, PA, USA }

\author[0000-0002-0173-6453]{N.~Fraija}
\affiliation{Instituto de Astronom\'{i}a, Universidad Nacional Autónoma de México, Ciudad de Mexico, Mexico }

%\author{A.~Galván-Gámez}
%\affiliation{Instituto de Astronom\'{i}a, Universidad Nacional Autónoma de México, Ciudad de Mexico, Mexico }

%\author{D.~Garcia}
%\affiliation{Instituto de F\'{i}sica, Universidad Nacional Autónoma de México, Ciudad de Mexico, Mexico }

\author[0000-0002-4188-5584]{J.A.~García-González}
\affiliation{Tecnologico de Monterrey, Escuela de Ingeniería y Ciencias, Ave. Eugenio Garza Sada 2501, Monterrey, N.L., 64849, Mexico }

\author[0000-0003-1122-4168]{F.~Garfias}
\affiliation{Instituto de Astronom\'{i}a, Universidad Nacional Autónoma de México, Ciudad de Mexico, Mexico }

%\author[0000-0001-9745-5738]{G.~Giacinti}
%\affiliation{Max-Planck Institute for Nuclear Physics, D-69117 Heidelberg, Germany}

\author[0000-0001-8493-2144]{H.~Goksu}
\affiliation{Max-Planck Institute for Nuclear Physics, D-69117 Heidelberg, Germany}

\author[0000-0002-5209-5641]{M.M.~González}
\affiliation{Instituto de Astronom\'{i}a, Universidad Nacional Autónoma de México, Ciudad de Mexico, Mexico }

\author[0000-0002-9790-1299]{J.A.~Goodman}
\affiliation{Department of Physics, University of Maryland, College Park, MD, USA }

\author[0000-0001-9844-2648]{J.P.~Harding}
\affiliation{Physics Division, Los Alamos National Laboratory, Los Alamos, NM, USA }

\author[0000-0002-2565-8365]{S.~Hernandez}
\affiliation{Instituto de F\'{i}sica, Universidad Nacional Autónoma de México, Ciudad de Mexico, Mexico }

\author[0000-0002-1031-7760]{J.~Hinton}
\affiliation{Max-Planck Institute for Nuclear Physics, D-69117 Heidelberg, Germany}

\author{B.~Hona}
\affiliation{Department of Physics and Astronomy, University of Utah, Salt Lake City, UT, USA }

\author[0000-0002-5447-1786]{D.~Huang}
\affiliation{Department of Physics, Michigan Technological University, Houghton, MI, USA }

\author[0000-0002-5527-7141]{F.~Hueyotl-Zahuantitla}
\affiliation{Universidad Autónoma de Chiapas, Tuxtla Gutiérrez, Chiapas, Mexico}

\author{P.~Hüntemeyer}
\affiliation{Department of Physics, Michigan Technological University, Houghton, MI, USA }

\author[0000-0001-5811-5167]{A.~Iriarte}
\affiliation{Instituto de Astronom\'{i}a, Universidad Nacional Autónoma de México, Ciudad de Mexico, Mexico }

\author[0000-0002-6738-9351]{A.~Jardin-Blicq}
\affiliation{Max-Planck Institute for Nuclear Physics, D-69117 Heidelberg, Germany}
\affiliation{Department of Physics, Faculty of Science, Chulalongkorn University, 254 Phayathai Road, Pathumwan, Bangkok 10330, Thailand}
\affiliation{National Astronomical Research Institute of Thailand (Public Organization), Don Kaeo, MaeRim, Chiang Mai 50180, Thailand}

\author[0000-0003-4467-3621]{V.~Joshi}
\affiliation{Friedrich-Alexander-Universit\"at Erlangen-N\"urnberg, Erlangen Centre for Astroparticle Physics, Erwin-Rommel-Stra\ss e 1, D-91058 Erlangen, Germany}

\author{S.~Kaufmann}
\affiliation{Universidad Politecnica de Pachuca, Pachuca, Hgo, Mexico }

\author[0000-0003-4785-0101]{D.~Kieda}
\affiliation{Department of Physics and Astronomy, University of Utah, Salt Lake City, UT, USA }

\author[0000-0002-2467-5673]{W.H.~Lee}
\affiliation{Instituto de Astronom\'{i}a, Universidad Nacional Autónoma de México, Ciudad de Mexico, Mexico }

\author[0000-0001-5516-4975]{H.~León Vargas}
\affiliation{Instituto de F\'{i}sica, Universidad Nacional Autónoma de México, Ciudad de Mexico, Mexico }

\author[0000-0003-2696-947X]{J.T.~Linnemann}
\affiliation{Department of Physics and Astronomy, Michigan State University, East Lansing, MI, USA }

\author[0000-0001-8825-3624]{A.L.~Longinotti}
\affiliation{Instituto Nacional de Astrof\'{i}sica, Óptica y Electrónica, Puebla, Mexico }

\author[0000-0003-2810-4867]{G.~Luis-Raya}
\affiliation{Universidad Politecnica de Pachuca, Pachuca, Hgo, Mexico }

\author[0000-0002-3882-9477]{R.~López-Coto}
\affiliation{INFN and Universita di Padova, via Marzolo 8, I-35131,Padova,Italy}

\author[0000-0001-8088-400X]{K.~Malone}
\affiliation{Space Science and Applications group, Los Alamos National Laboratory, Los Alamos, NM, USA }

\author[0000-0001-9077-4058]{V.~Marandon}
\affiliation{Max-Planck Institute for Nuclear Physics, D-69117 Heidelberg, Germany}

\author[0000-0001-9052-856X]{O.~Martinez}
\affiliation{Facultad de Ciencias F\'{i}sico Matemáticas, Benemérita Universidad Autónoma de Puebla, Puebla, Mexico }

%\author[0000-0001-9035-1290]{I.~Martinez-Castellanos}
%\affiliation{Department of Physics, University of Maryland, College Park, MD, USA }

\author[0000-0002-2824-3544]{J.~Martínez-Castro}
\affiliation{Centro de Investigaci\'on en Computaci\'on, Instituto Polit\'ecnico Nacional, M\'exico City, Mexico.}

\author[0000-0002-2610-863X]{J.A.~Matthews}
\affiliation{Dept of Physics and Astronomy, University of New Mexico, Albuquerque, NM, USA }

\author[0000-0002-8390-9011]{P.~Miranda-Romagnoli}
\affiliation{Universidad Autónoma del Estado de Hidalgo, Pachuca, Mexico }

\author[0000-0001-9361-0147]{J.A.~Morales-Soto}
\affiliation{Universidad Michoacana de San Nicolás de Hidalgo, Morelia, Mexico }

\author[0000-0002-1114-2640]{E.~Moreno}
\affiliation{Facultad de Ciencias F\'{i}sico Matemáticas, Benemérita Universidad Autónoma de Puebla, Puebla, Mexico }

\author[0000-0002-7675-4656]{M.~Mostafá}
\affiliation{Department of Physics, Pennsylvania State University, University Park, PA, USA }

\author[0000-0003-0587-4324]{A.~Nayerhoda}
\affiliation{Institute of Nuclear Physics Polish Academy of Sciences, PL-31342 IFJ-PAN, Krakow, Poland }

\author[0000-0003-1059-8731]{L.~Nellen}
\affiliation{Instituto de Ciencias Nucleares, Universidad Nacional Autónoma de Mexico, Ciudad de Mexico, Mexico }

\author[0000-0001-9428-7572]{M.~Newbold}
\affiliation{Department of Physics and Astronomy, University of Utah, Salt Lake City, UT, USA }

\author[0000-0002-6859-3944]{M.U.~Nisa}
\affiliation{Department of Physics and Astronomy, Michigan State University, East Lansing, MI, USA }

\author[0000-0001-7099-108X]{R.~Noriega-Papaqui}
\affiliation{Universidad Autónoma del Estado de Hidalgo, Pachuca, Mexico }

\author[0000-0002-9105-0518]{L.~Olivera-Nieto}
\affiliation{Max-Planck Institute for Nuclear Physics, D-69117 Heidelberg, Germany}

\author[0000-0002-5448-7577]{N.~Omodei}
\affiliation{Department of Physics, Stanford University: Stanford, CA 94305–4060, USA}

\author{A.~Peisker}
\affiliation{Department of Physics and Astronomy, Michigan State University, East Lansing, MI, USA }

\author[0000-0002-8774-8147]{Y.~Pérez Araujo}
\affiliation{Instituto de Astronom\'{i}a, Universidad Nacional Autónoma de México, Ciudad de Mexico, Mexico }

\author[0000-0001-5998-4938]{E.G.~Pérez-Pérez}
\affiliation{Universidad Politecnica de Pachuca, Pachuca, Hgo, Mexico }

\author[0000-0002-6524-9769]{C.D.~Rho}
\affiliation{University of Seoul, Seoul, Republic Of Korea}

\author[0000-0003-1327-0838]{D.~Rosa-González}
\affiliation{Instituto Nacional de Astrof\'{i}sica, Óptica y Electrónica, Puebla, Mexico }

\author[0000-0001-6939-7825]{E.~Ruiz-Velasco}
\affiliation{Max-Planck Institute for Nuclear Physics, D-69117 Heidelberg, Germany}

\author{H.~Salazar}
\affiliation{Facultad de Ciencias F\'{i}sico Matemáticas, Benemérita Universidad Autónoma de Puebla, Puebla, Mexico}

\author{D.~Salazar-Gallegos}
\affiliation{Department of Physics and Astronomy, Michigan State University, East Lansing, MI, USA }

\author[0000-0002-8610-8703]{F.~Salesa Greus}
\affiliation{Institute of Nuclear Physics Polish Academy of Sciences, PL-31342 IFJ-PAN, Krakow, Poland }
\affiliation{Instituto de F\'{i}sica Corpuscular, CSIC, Universitat de Val\`{e}ncia, E-46980, Paterna, Valencia, Spain}

\author[0000-0001-6079-2722]{A.~Sandoval}
\affiliation{Instituto de F\'{i}sica, Universidad Nacional Autónoma de México, Ciudad de Mexico, Mexico }

\author[0000-0001-8644-4734]{M.~Schneider}
\affiliation{Department of Physics, University of Maryland, College Park, MD, USA }

%\author[0000-0002-8999-9249]{H.~Schoorlemmer}
%\affiliation{Max-Planck Institute for Nuclear Physics, D-69117 Heidelberg, Germany}

\author{J.~Serna-Franco}
\affiliation{Instituto de F\'{i}sica, Universidad Nacional Autónoma de México, Ciudad de Mexico, Mexico }

\author[0000-0002-1012-0431]{A.J.~Smith}
\affiliation{Department of Physics, University of Maryland, College Park, MD, USA }

\author{Y.~Son}
\affiliation{University of Seoul, Seoul, Republic Of Korea}

\author[0000-0002-1492-0380]{R.W.~Springer}
\affiliation{Department of Physics and Astronomy, University of Utah, Salt Lake City, UT, USA }

%\author[0000-0002-8516-6469]{P.~Surajbali}
%\affiliation{Max-Planck Institute for Nuclear Physics, D-69117 Heidelberg, Germany}

\author{O.~Tibolla}
\affiliation{Universidad Politecnica de Pachuca, Pachuca, Hgo, Mexico }

\author[0000-0001-9725-1479]{K.~Tollefson}
\affiliation{Department of Physics and Astronomy, Michigan State University, East Lansing, MI, USA }

\author[0000-0002-1689-3945]{I.~Torres}
\affiliation{Instituto Nacional de Astrof\'{i}sica, Óptica y Electrónica, Puebla, Mexico }

\author[0000-0002-7102-3352]{R.~Torres-Escobedo}
\affiliation{Tsung-Dao Lee Institute \&{} School of Physics and Astronomy, Shanghai Jiao Tong University, Shanghai, People's Republic of China }

\author[0000-0003-1068-6707]{R.~Turner}
\affiliation{Department of Physics, Michigan Technological University, Houghton, MI, USA}

\author[0000-0002-2748-2527]{F.~Ureña-Mena}
\affiliation{Instituto Nacional de Astrof\'{i}sica, Óptica y Electrónica, Puebla, Mexico }

\author[0000-0001-6876-2800]{L.~Villaseñor}
\affiliation{Facultad de Ciencias F\'{i}sico Matemáticas, Benemérita Universidad Autónoma de Puebla, Puebla, Mexico }

\author[0000-0001-6798-353X]{X.~Wang}
\affiliation{Department of Physics, Michigan Technological University, Houghton, MI, USA }

%\author{T.~Weisgarber}
%\affiliation{Department of Physics, University of Wisconsin-Madison, Madison, WI, USA}

\author[0000-0002-6941-1073]{F.~Werner}
\affiliation{Max-Planck Institute for Nuclear Physics, D-69117 Heidelberg, Germany}

\author[0000-0002-6623-0277]{E.~Willox}
\affiliation{Department of Physics, University of Maryland, College Park, MD, USA }

\author[0000-0003-0513-3841]{H.~Zhou}
\affiliation{Tsung-Dao Lee Institute \&{} School of Physics and Astronomy, Shanghai Jiao Tong University, Shanghai, People's Republic of China }

\collaboration{95}{(HAWC collaboration)}

%% Note that the \and command from previous versions of AASTeX is now
%% depreciated in this version as it is no longer necessary. AASTeX 
%% automatically takes care of all commas and "and"s between authors names.

%% AASTeX 6.31 has the new \collaboration and \nocollaboration commands to
%% provide the collaboration status of a group of authors. These commands 
%% can be used either before or after the list of corresponding authors. The
%% argument for \collaboration is the collaboration identifier. Authors are
%% encouraged to surround collaboration identifiers with ()s. The 
%% \nocollaboration command takes no argument and exists to indicate that
%% the nearby authors are not part of surrounding collaborations.

%% Mark off the abstract in the ``abstract'' environment. 
\begin{abstract}

The latest High Altitude Water Cherenkov (HAWC) point-like source catalog up to ~56 TeV reported the detection of two sources in the region of the Galactic plane at galactic longitude 52\textdegree \ $< \ell <$ 55\textdegree, 3HWC~J1930+188 and 3HWC~J1928+178. The first one is associated with a known TeV source, the supernova remnant SNR~G054.1+00.3. It was discovered by one of the currently operating Imaging Atmospheric Cherenkov Telescope (IACT), the Very Energetic Radiation Imaging Telescope Array System (VERITAS), detected by the High Energy Stereoscopic System (H.E.S.S.), and identified as a composite SNR. 
However, the source 3HWC J1928+178, discovered by HAWC and coincident with the pulsar PSR~J1928+1746, was not detected by any IACT despite their long exposure on the region, until a recent new analysis of H.E.S.S. data was able to confirm it. 
Moreover, no X-ray counterpart has been detected from this pulsar. 
We present a multicomponent fit of this region using the latest HAWC data. This reveals an additional new source, HAWC~J1932+192, which is potentially associated with the pulsar PSR~J1932+1916, whose $\gamma$-ray emission could come from the acceleration of particles in its pulsar wind nebula. In the case of 3HWC~J1928+178, several possible explanations are explored, in a attempt to unveil the origins of the very-high-energy $\gamma$-ray emission.

\end{abstract}

%% Keywords should appear after the \end{abstract} command. 
%% The AAS Journals now uses Unified Astronomy Thesaurus concepts:
%% https://astrothesaurus.org
%% You will be asked to selected these concepts during the submission process
%% but this old "keyword" functionality is maintained in case authors want
%% to include these concepts in their preprints.

%% From the front matter, we move on to the body of the paper.
%% Sections are demarcated by \section and \subsection, respectively.
%% Observe the use of the LaTeX \label
%% command after the \subsection to give a symbolic KEY to the
%% subsection for cross-referencing in a \ref command.
%% You can use LaTeX's \ref and \label commands to keep track of
%% cross-references to sections, equations, tables, and figures.
%% That way, if you change the order of any elements, LaTeX will
%% automatically renumber them.
%%
%% We recommend that authors also use the natbib \citep
%% and \citet commands to identify citations.  The citations are
%% tied to the reference list via symbolic KEYs. The KEY corresponds
%% to the KEY in the \bibitem in the reference list below. 

\section{Introduction} 
\label{Introduction}

The large majority of the TeV $\gamma$-ray sources detected so far, mainly thanks to surveys like the the High Energy Stereoscopic System (H.E.S.S.) Galactic Plane Survey (HGPS; ~\citealt{HGPS}), are located in the Galactic plane, and most of them remain unidentified. More generally, the origin of the observed $\gamma$-ray emission is often uncertain. Indeed, while the Galactic plane is the best place to look for TeV $\gamma$-ray sources, it is a quite complex region in itself: the proximity of Galactic plane sources leads to source confusion, and large-scale diffuse emission needs to be taken into account. However, the diffuse emission is poorly understood and not very well modeled, partially due to our lack of knowledge about the gas distribution and the distribution of unresolved sources. In addition, the magnetic field structures can be quite complex and difficult to assess. The Galactic plane is also the place for star formation, involving giant molecular clouds (GMCs) that imply different kinds of interactions, shocks, propagation and diffusion processes~\citep{Molecular_cloud_SFR, CR_propagation, galactic_CR}. The modeling of complex regions and the detailed morphological and spectral analysis of individual sources are crucial for testing different scenarios and obtaining a better understanding of the origin of the observed $\gamma$-ray emission. 
The very-high-energy (VHE; $E>100$ GeV) $\gamma$-ray emission of the sources 3HWC~J1928+178 and 3HWC~J1930+188, reported in the third High Altitude Water Cherenkov (HAWC) catalog~\citep{3HWC_catalog} at the galactic coordinates (52\textdegree93,~0\textdegree20) and (54\textdegree03,~0\textdegree32) respectively, and the new source HAWC~J1928+192 located at (54\textdegree69,~0\textdegree20), are the focus of this paper. Because of their possible association with pulsars, a classical pulsar wind nebula (PWN) scenario is studied. However, a molecular cloud in the vicinity of 3HWC~J1928+178 makes it a perfect candidate for studying the possible interaction of charged particles with the components of the cloud. After presenting a multiwavelength picture of the region in section~\ref{Multi-wavelength observations of the region}, and an overview of the HAWC data in section~\ref{HAWC observations}, we present the multicomponent modeling of the region in section~\ref{Method} and the results of the fit using a maximum likelihood approach in section~\ref{Results}. Section~\ref{Origin} is dedicated to an assessment of different hypotheses regarding the origin of the $\gamma$-ray emission of 3HWC~J1928+178. In particular, a scenario involving Inverse Compton (IC) scattering is considered, as well as possible interaction with a nearby molecular cloud.
The conclusion is drawn in Section~\ref{Conclusion}. \\
\newpage

%%%%%%%%%%%%%%%%%%%%%%%
%\section{Observations}
%\label{Observation}

\section{Multiwavelength picture of the region}
\label{Multi-wavelength observations of the region}

\begin{figure}[ht!]
    \centering
    \includegraphics[width=0.9\linewidth]{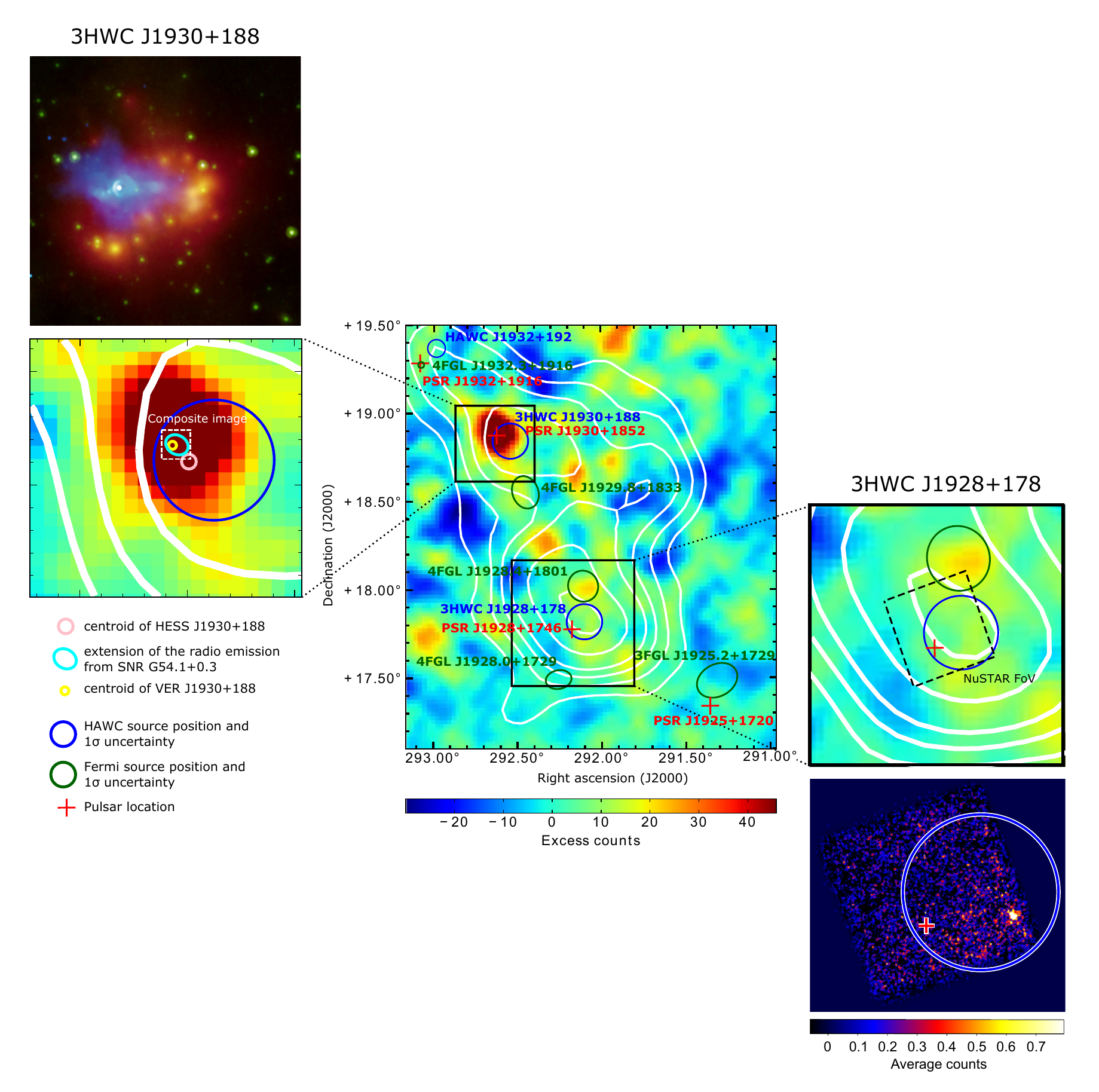}
    \caption{Multiwavelength view of the region surrounding 3HWC J1928+177. The middle map is the VERITAS excess map of the region, adapted from~\citet{Veritas_Fermi_2HWCsources}. Superimposed are the locations and the 1$\sigma$ uncertainties on the locations of the HAWC sources (blue circles) and the Fermi 4FGL sources (green circles), as well as the locations of the pulsars (red crosses). The white contours are HAWC significance contours for 5$\sigma$, 6$\sigma$, 7$\sigma$, 8$\sigma$, 10$\sigma$ and 12$\sigma$ for 1523 days of data. 
    The top source, 3HWC~J1930+188, is detailed in the zoomed-in view on the left-hand side. The locations of the counterparts detected by VERITAS (yellow) and H.E.S.S. (pink) are represented. The extension of the radio emission is also shown (cyan). The dashed white box represents the size of the composite image at the top (3$\arcmin$ - 0\textdegree05). It depicts the X-ray emission of the pulsar (the bright white star) and the PWN detected by \textit{Chandra} (blue - NASA/CXC/SAO/T.Temim et al.), as well as the IR emission detected by \textit{Spitzer} (green is 8$\mu$m and red is 24$\mu$m - NASA/JPL-Caltech), revealing the dusty remains of a collapsed star. The bottom source 3HWC~J1928+178 is detailed in the zoomed-in image on the right-hand side. The dashed black box represents the NuSTAR background-subtracted map shown at the bottom (adapted from~\citet{J1928_dark_accelerator}). The bright source to the bottom right is CXO J192812.0+174712.
    }
    \label{Veritas_Chandra_NuSTAR}
\end{figure}

\textbf{3HWC~J1930+188} is associated with the $\gamma$-ray emission of the PWN in the supernova remnant SNR~G54.1+0.3, located at 6.2~kpc~\citep{PWNG54_CO}. Studies of the X-ray emission using \textit{XMM-Newton} and \textit{Suzaku} data have inferred that the SNR~G54.1+0.3 would be $\sim$2000~years old ~\citep{G54_age}. It was first detected with 6.8$\sigma$ significance by the Very Energetic Radiation Imaging Telescope Array System (VERITAS) in 2010, with a total observation time of 36.6~hr~\citep{VERITAS_J1930}, and identified as the point-like source VER~J1930+188. With 16 additional hours of observation in 2015-2016~\citep{Veritas_Fermi_2HWCsources}, there is now a total exposure time of 46 hr from VERITAS on this region. Figure~\ref{Veritas_Chandra_NuSTAR} shows the latest VERITAS excess map of the region, zooming in on each HAWC source~\citep{Veritas_Fermi_2HWCsources}. It was shown that the centroid of the HAWC detection agrees with the VERITAS centroid position. However, the spectral index of the simple power law found for the HAWC source, $-2.74 \pm 0.12_{\mbox{\scriptsize stat}}$, is softer than that measured by VERITAS, $-2.18 \pm 0.2_{\mbox{\scriptsize stat}}$. Moreover, the differential flux at 7 TeV measured by HAWC is $(9.8 \pm 1.5) \times 10^{-15}$~TeV$^{-1}$~cm$^{-2}$~s$^{-1}$ while the differential flux at 1 TeV measured by VERITAS is $(6.6 \pm 1.3) \times 10^{-13}$~TeV$^{-1}$~cm$^{-2}$~s$^{-1}$. Extrapolating the HAWC spectrum to the VERITAS energy range gives an integrated flux seven times larger than the VERITAS flux, although it is still within the 2$\sigma$ statistical uncertainties of the VERITAS measurement. 
The H.E.S.S. collaboration has also reported the detection of this source in the HGPS~\citep{HGPS} catalog and referenced it as a composite SNR, as it was not possible to distinguish the origin of the emission between the shell and the PWN. 
At the center of the PWN, the pulsar PSR~J1930+1852 was discovered in 2002 by the Arecibo radio telescope, with a period of 136~ms ~\citep{Discovery_PSRJ1930}. With a derived spin-down power of $\dot{E}~=~1.2 \times 10^{37}$ erg~s$^{-1}$ and a characteristic age of $\sim$2900~yr~\citep{Discovery_PSRJ1930}, it is amongst the youngest and most energetic known pulsars. 
Observations of the X-ray emission by the \textit{Chandra} X-ray observatory over 290.77 ks  reveal the pulsar and the PWN~\citep{G54_Chandra_Spiter}. In addition, IR observations by the \textit{Spitzer} space telescope~\citep{G54_Chandra_Spiter} and the \textit{Herschel} space observatory~\citep{G54_Hershel} show a shell of gas and dust, debris from the supernova explosion. 
The shell contains compact IR sources arranged in a ringlike structure. These may be young stellar objects, whose formation would have been triggered by the wind of the progenitor star~\citep{G54_IRshell}. They could also be ejecta dust heated by early-type stars belonging to the stellar cluster in which the star exploded~\citep{G54_Chandra_Spiter}. Both \textit{Chandra} X-ray and \textit{Spitzer} IR images are visible in the composite image in the left part of Figure \ref{Veritas_Chandra_NuSTAR}. A morphological association with a molecular cloud detected from CO observations has been suggested~\citep{PWNG54_CO}, but no evidence for interaction with this cloud was found~\citep{CO_G54}. A $^{12}$CO map (rotation emission line $J=1\rightarrow0$ at 115~GHz;~\citealt{COsurvey}) and a radio map from the GaLactic and Extragalactic All-sky MWA (GLEAM) survey~\citep{GLEAMsurvey} are shown in the right panel of Figure~\ref{MWL_1523d_CO_radio} where HAWC significance contours have been superimposed. This source will be referred to as J1930 hereafter. All the details relating to this source are summarized in Table~\ref{J1930_caracteristics} in the Appendix~\ref{Multi-wavelength information}. \\

\textbf{3HWC J1928+178} is located about one degree away from 3HWC~J1930+188. It was not detected by any Imaging Atmospheric Cherenkov Telescope (IACT), despite the 46 and 36 hr of observations by VERITAS and H.E.S.S., respectively, until H.E.S.S. could confirm a detection with significant emission above 5$\sigma$ using a new analysis method more appropriate for extended sources~\citep{HAWC-HESS_GP_ApJ}. It is detected by HAWC with more than 12$\sigma$. 
It is likely associated with the pulsar PSR~J1928+1746, located 0\textdegree03 away from the 3HWC source location, one of the pulsars discovered at radio wavelength in 2006 in a long-term pulsar survey of the Galactic plane using the Arecibo L-band Feed Array (ALFA;~\citealt{Discovery_PSRJ1928}). It is described as a young isolated pulsar with a period of 68.7~ms, a spin-down power of $\dot{E}$~=~$1.6\times10^{36}$~erg~s$^{-1}$ and a characteristic age of 82~kyr. The distance to it is estimated to be 4.3~kpc~\citep{ymw17}. No detections in X-ray by \textit{Chandra} or NuSTAR have been reported for this pulsar, as depicted by the bottom right-hand part of Figure \ref{Veritas_Chandra_NuSTAR}. However, the variable X-ray source CXO~J192812.0+174712 is found within the 3HWC source position uncertainties. The association with the 3HWC source has been studied by \citet{J1928_dark_accelerator} in the case of a binary system, although no variability has been seen at TeV energies. Finally, the unidentified Fermi source 4FGL~J1928.4+1801 is located 0\textdegree1 away from the 3HWC source. 
This source will be referred to as J1928 hereafter. All the details relating to this source are summarized in Table~\ref{J1928_caracteristics} in the Appendix~\ref{Multi-wavelength information}.\\

\textbf{HAWC J1932+192} is spatially coincident with the pulsar PSR~J1932+1916, discovered by the \textit{Fermi}-LAT in 2013~\citep{Discovery_PSRJ1932} and classified as radio-quiet. It has a period of 208~ms, a spin-down power $\dot{E}$~=~$4.07\times10^{35}$~erg~s$^{-1}$ and a characteristic age of 35.4~kyr. It has also been observed in X-ray by \textit{Suzaku} and by the \textit{Swift} X-ray telescope, and an extended X-ray emission has been reported ~\citep{J1932_Xrays}. In that study, the emission was modeled with two Gaussians: a narrow one with a FWHM~$\leq0\arcmin5$ which could be associated with the pulsar, and a broad one with a FWHM of~$\sim$4$\arcmin$5, which could be interpreted as the PWN emission.
Using these observations, its distance is estimated as being between 2 and 6~kpc~\citep{J1932_Xrays}. This emission is located near the edge of the SNR~G54.4-0.3. It is clearly visible on the radio map, in the lower right-hand panel of Figure~\ref{MWL_1523d_CO_radio} in the shape of a circular feature with the pulsar on the edge. 
Moreover, a CO structure was reported to be in morphological coincidence with the radio emission, with an evidence for the interaction of the SNR with the surrounding CO shell~\citep{G54.4_CO}. 
For this SNR, the distance has been estimated as being 6.6~kpc~\citep{G54.4}. This source will be referred to as J1932 hereafter. All the details relating to this source are summarized in Table~\ref{J1932_caracteristics} in the Appendix~\ref{Multi-wavelength information}.

%\newpage

%\newpage 

\section{HAWC observations}
\label{HAWC observations}

\begin{figure}[ht!]
    \centering
    \includegraphics[width=1.0\linewidth]{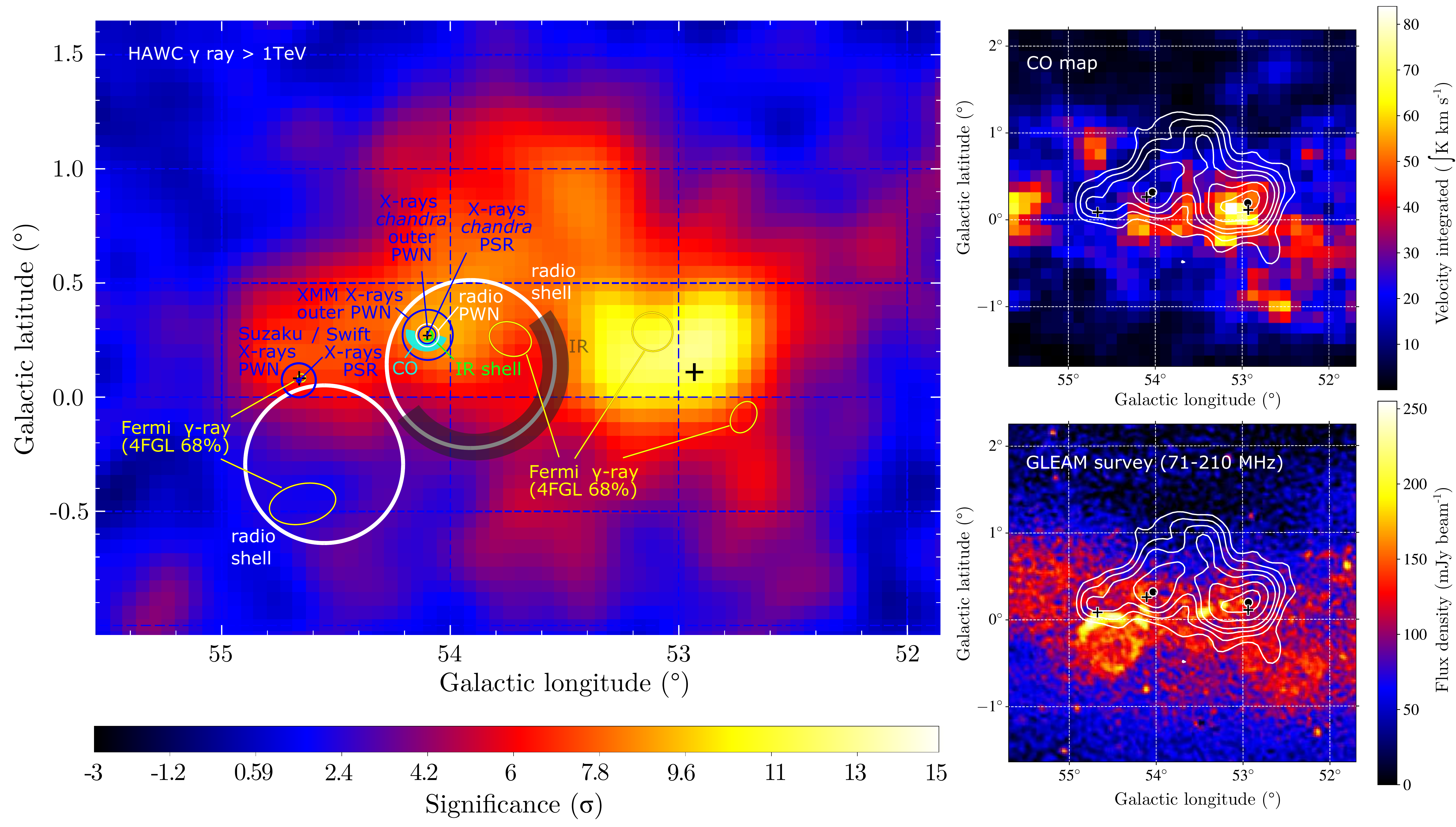}
    \caption{Left: X-ray, radio, IR and GeV $\gamma$-ray emission superimposed on the HAWC significance map for 1523 days, using analysis bins 4 to 9. Top right: velocity-integrated CO map~\citep{COsurvey}. Bottom right: 71-210 MHz radio map from the GLEAM survey ~\citep{GLEAMsurvey}. Superimposed are the HAWC contours for 5$\sigma$, 6$\sigma$, 7$\sigma$, 8$\sigma$, 10$\sigma$ and 12$\sigma$. The locations of the HAWC sources are represented by the black dots. The positions of the pulsars PSR~J1930+1852, PSR~J1932+1916, and PSR~J1928+1746 are (292\textdegree63, 18\textdegree87), (293\textdegree08, 19\textdegree28) and (292\textdegree17,  17\textdegree77), respectively, according to the ATNF catalog~\citep{ATNFcatalog}, and are represented by the black crosses.
    }
    \label{MWL_1523d_CO_radio}
\end{figure}

HAWC is an array of 300 water tanks covering an area of 22,000 m$^2$, each instrumented with four photomultiplier tubes. 
The $\gamma$-ray-like events are classified with respect to the fraction of the array that was triggered. 
They are assigned to one of the nine analysis bins, according to the definition in~\cite{HAWC_crab}, from analysis bin 1, gathering events triggering 7\% to 10\% of the array, to analysis bin 9, for events hitting 84\% to 100\% of the array. Low-energy events that trigger only a small fraction of the array are likely to be found in the low analysis bins, while the highest-energy events triggering most of the array will be found in the higher analysis bins. This analysis is restricted to bins 4 to 9, as a good compromise between reasonable performance at TeV energies and enough statistics. Indeed, the greater the fraction of the array that was hit, the more information is available and the lower the uncertainties on the reconstructed parameters. In particular, the $\gamma$/hadron separation improves with the increase in the analysis bins, reaching an efficiency of $1\times10^{-2}$ to $1\times10^{-3}$ for events in analysis bins 4 to 9.  
The HAWC significance map of the region for 1523 days of data, produced with the reconstruction Pass 4, under the hypothesis of a point-like source and a spectral index of $-2.5$, is shown in Figure~\ref{MWL_1523d_CO_radio}. 
Two sources, 3HWC~J1930+188 and 3HWC~J1928+178, are reported in the 3HWC~catalog~\citep{3HWC_catalog}. 

%%%%%%%%%%%%%%%%%%%%%%%%%%%%%%%%%%%%%%%%%%%%%%%%%%%%%%%%%%%%%%%%
\section{Method: the modeling of the region and fit of the HAWC data} 
\label{Method}

Modeling this region is not trivial, because it requires disentangling the different sources of emission. An attempt to represent this complex region with several components is described here. For each component, the parameters are fitted simultaneously using the Multi-Mission Maximum Likelihood framework\footnote{The documentation is available at https://threeml.readthedocs.io/en/stable/ and the code at https://github.com/threeML/threeML} (3ML;~\citealt{3ML}) and the HAWC HAL plugin\footnote{The documentation and code are available at https://github.com/threeML/hawc\_hal}~\citep{ICRC_HAL}. This is based on a maximum likelihood approach, in which a model representing a particular region of the sky, here made of several components, is convolved with the instrument response and compared to the corresponding experimental data.
%\newpage
An initial model is defined for the region based on our current knowledge: 
\begin{itemize}
 \item VER J1930+188 and HESS J1930+188 are point-like sources associated with the PWN surrounding the pulsar PSR~J1930+1852. Hence, the source 3HWC~J1930+188 is defined as a point-like source initialized at the location of the pulsar (292\textdegree63, 18\textdegree87) . This component will be used to model J1930. 
 
 \item The source 3HWC~J1928+177 is represented by a symmetric Gaussian with the initial location at the position of the pulsar PSR~J1928+1746 (292\textdegree18, 17\textdegree77) and an initial size of $\sigma$ = 0\textdegree1. This component will be used to model J1928.
\end{itemize}
These are the two components of the initial model, visible in the panel (a) of Figure~\ref{summary_a}. There is no component for the galactic diffuse emission. The positions of the two components and the size of the extended component are left free. Their spectra are assumed to follow a simple power law with free index initialized at $-2.5$ and free differential flux at 10~TeV initialized at $1.0\times10^{-14}$~TeV$^{-1}$~cm$^{-2}$~s$^{-1}$. The fit is performed in an iterative process, starting by fitting the initial model to the data. For each component, a test statistic (TS) is computed, which compares the likelihood that a source is present against the hypothesis that there is no source but only background fluctuations: 
\begin{equation}
 \mbox{TS} = 2~\mbox{ln}\frac{\mathcal{L}(\mbox{source model})}{\mathcal{L}(\mbox{no source})}.
\end{equation}
If a remaining excess is found in the residual map, a point-like component with a power-law spectrum is added at the location of the excess, and the fit is performed again with the position and spectral parameters being free. This new component is kept if it significantly improves the fit, by $\Delta \mbox{TS} = 25 $. 

%%%%%%%%%%%%%%%%%%%%%%%%%%%%%%%%%%%%%%%%%%%%%%%%%%%%%%%%%%%%%%%%
\section{Results} 
\label{Results}
\subsection{Results of the fit}
\begin{figure}[ht!]
    \centering
    \includegraphics[width=0.93\linewidth]{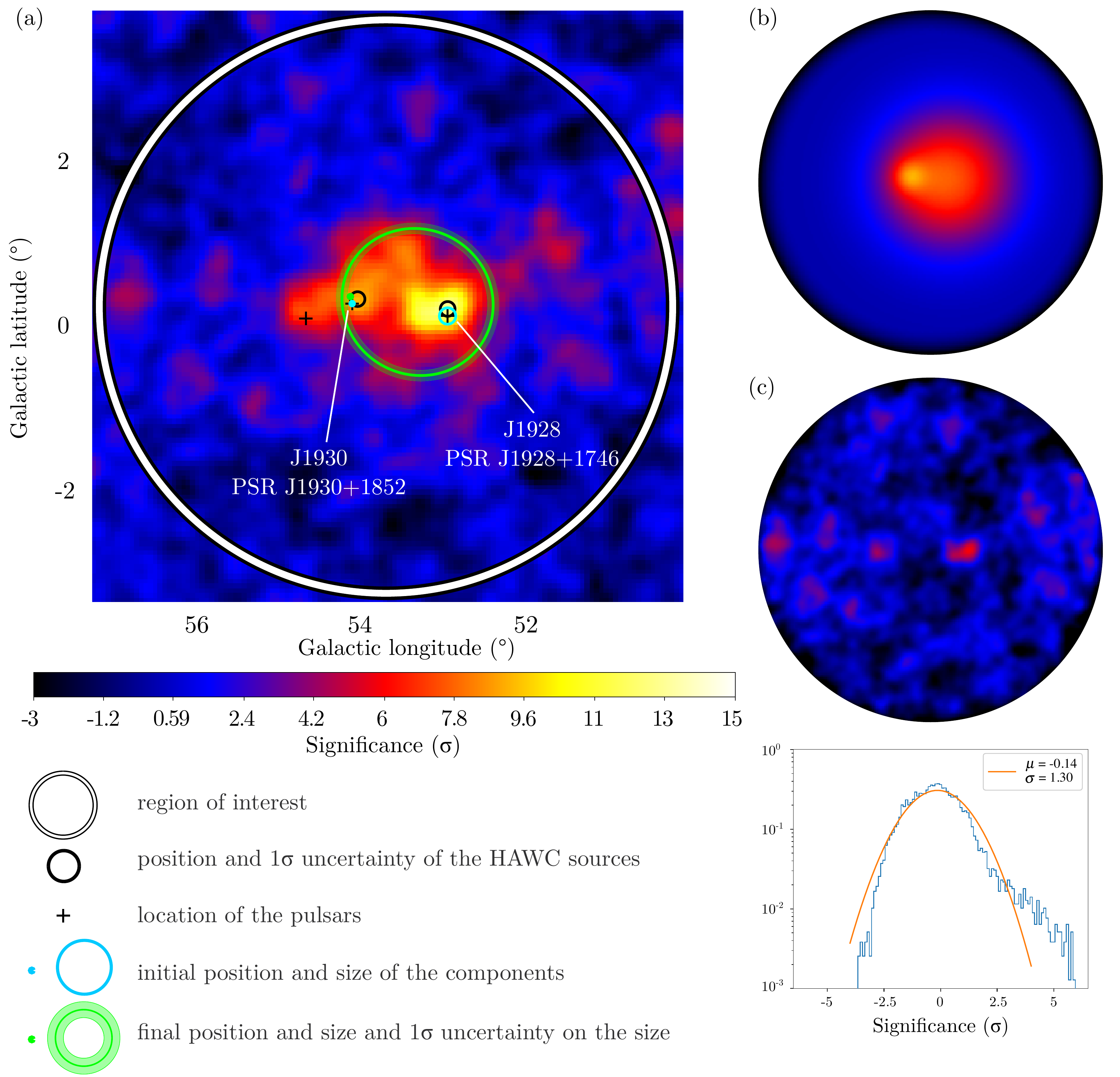}
    \caption{ The significance map (a) indicates the region of interest (ROI) of radius 3\textdegree5 (the white circle) and the two components for J1928 and J1930. The blue/green dot and circle show the initial/fitted position and size. The width of the green circle represents the 1$\sigma$ uncertainty on the size of the Gaussian. 
    Map (b) is the significance map of the model in the ROI. 
    Map (c) is the significance map of the residuals in the ROI and the significance distribution in the inner 2\textdegree \ radius region, with a Gaussian fit. The color scale holds for all maps.
    }
    \label{summary_a}
\end{figure}

\begin{figure}[ht!]
    \centering
    \includegraphics[width=0.93\linewidth]{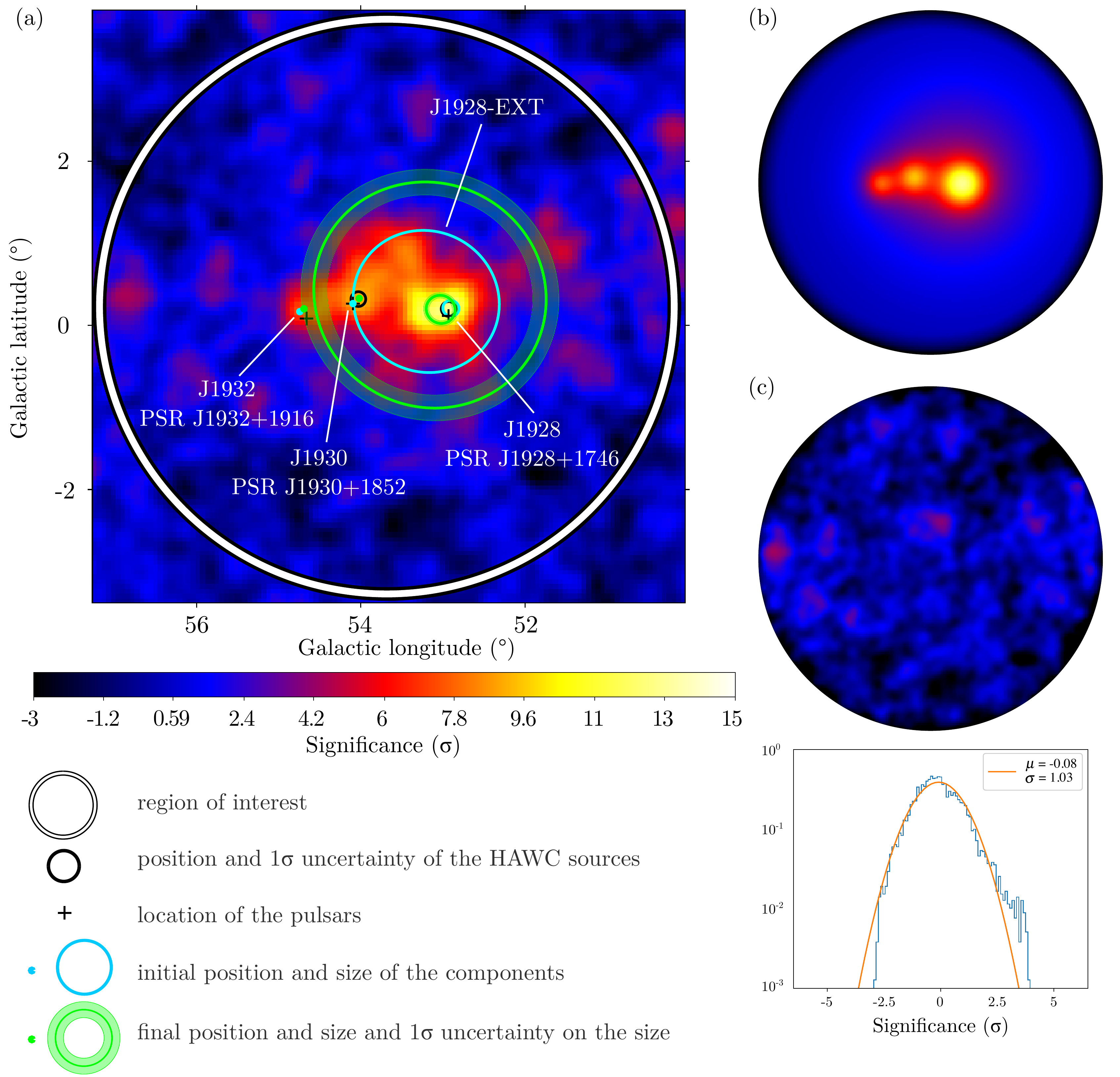}
    \caption{ The significance map (a) shows the region of interest (ROI) of radius 3.5\textdegree \ (the white circle) and the four components for J1928, J1930, and J1932, as well as the additional extended source J1928-EXT. The blue/green dots and circles show their initial/fitted positions and sizes. The width of the green circle represent the 1$\sigma$ uncertainty on the size of the Gaussian. 
    Map (b) is the significance map of the model in the ROI. 
    Map (c) is the significance map of the residuals in the ROI and the significance distribution in the inner 2\textdegree \ radius region, with a Gaussian fit. The color scale holds for all maps. 
    }
    \label{summary_b}
\end{figure}

\begin{figure}[ht!]
    \centering
    \includegraphics[width=0.8\linewidth]{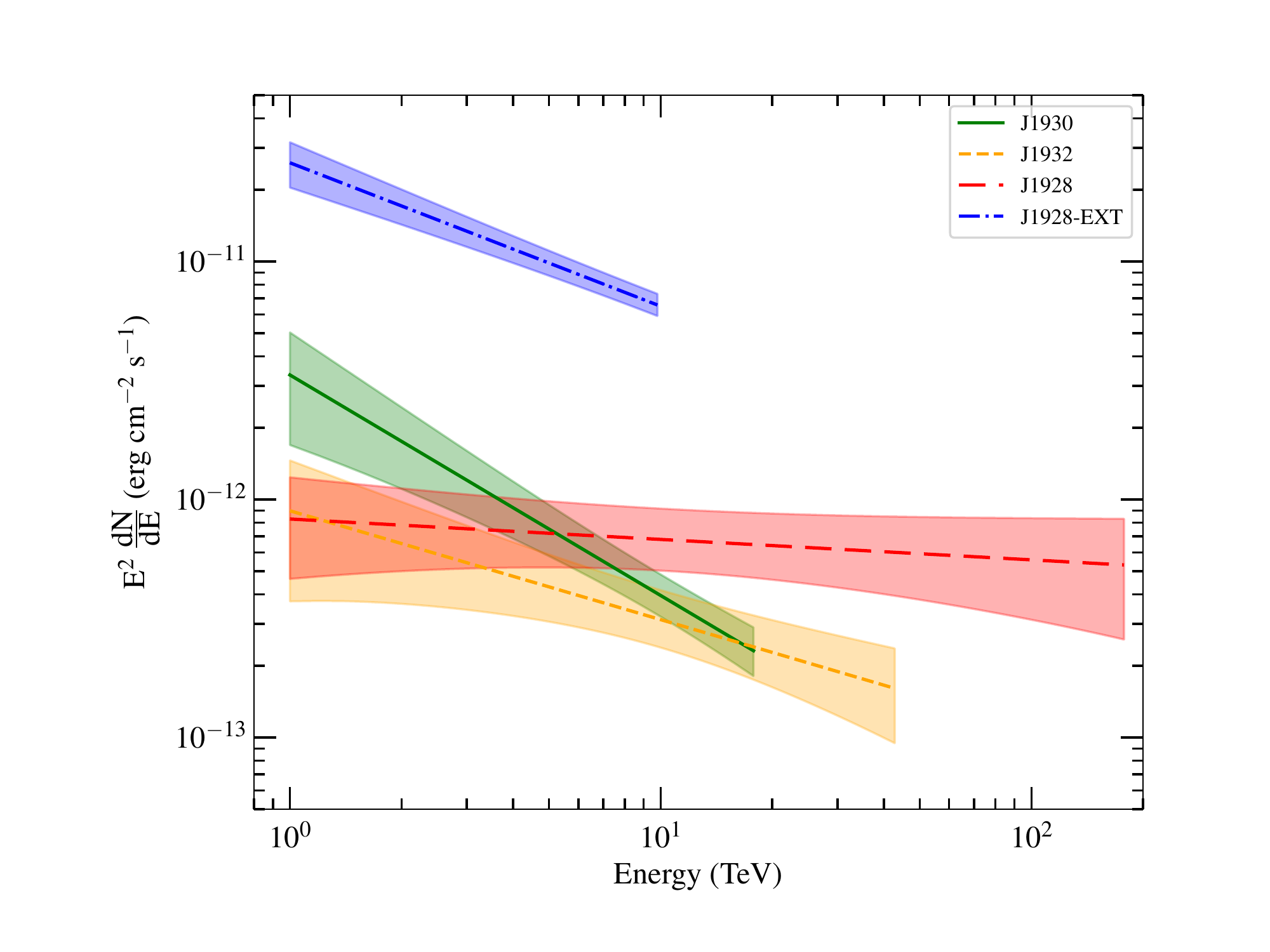}
    \caption{ Spectral energy distribution of the four components of the best-fit model. The spectral parameters are given in Table~\ref{model_parameters}. The shaded bands are the 1$\sigma$ statistical uncertainties.
    }
    \label{all_spec-PL}
\end{figure}

Figure~\ref{summary_a}(a) shows the HAWC significance map of the region for a point-like source hypothesis and assuming a power-law spectrum with an index of $-2.5$, which are the standard parameters used to produce HAWC maps~\citep{3HWC_catalog}.
Superimposed in blue and green are the initial and fitted positions of the two components previously described in section~\ref{Method}. The width of the green circle represents the 1$\sigma$ uncertainty on the size of the Gaussian. 
The fitted model is displayed in Figure~\ref{summary_a}(b) and the residual map and its distribution in Figure~\ref{summary_a}(c). The orange line is a fit to the distribution with a gaussian function. 
After this first iteration, excesses of $4\sigma$ and $6\sigma$ significance are found in the residual map near the pulsar PSR~J1932+1916 and at the location of J1928, respectively. To account for this, two components are added to the model at the locations of the excesses: 
\begin{itemize}
 \item A point-like source is initialized at ($\mbox{R. A.}=293^{\circ}07$, $\mbox{decl.}=19^{\circ}40$), near PSR~J1932+1916, with a simple power law as spectral model. This component will be simply called J1932.
 \item An extended source is initialized at ($\mbox{R. A.}=292^{\circ}08$, $\mbox{decl.}=17^{\circ}79$), with initial size $\sigma$~=~0\textdegree1, with a simple power law as a spectral model. This is the new component for J1928. 
\end{itemize}
The previous extended component from the initial model will now be called J1928-EXT, and it is given as the initial position and size the output from the first fit. The position, size, index, and flux normalization are again set free. A fit is performed again with the four components. The outputs of the second fit are summarized in Table~\ref{model_parameters}. The corresponding maps are displayed in Figure~\ref{summary_b}, using the same color code as in Figure~\ref{summary_a}. The spectra for the four components are shown in Figure~\ref{all_spec-PL}. The lower edges of the spectra are fixed to 1~TeV, as the median energy of analysis bin 4 minus an error of 1$\sigma$. To determine the upper edges, individual fits are performed for each of the four components using a power law with an exponential cutoff for that component only, with the amplitude and cutoff energy as the only free parameters, and the three other components being modeled by a simple power law with all parameters fixed. Then, the cutoff energy is fixed as well, so that only the flux normalization remains as a free parameter, and the cutoff energy is set to decreasing values until $\Delta TS=2$. In Table~\ref{model_parameters}, all of the fitted parameters are given with the statistical and systematic uncertainties. To calculate the systematic uncertainties, the same fit was performed again using different instrument response files.

The component representing J1928 is found to have a size of $\sigma~=~0^{\circ}18
\pm 0^{\circ}04_{\mbox{\scriptsize stat}}$ (39\% containment), while the other extended source, J1928-EXT, has a size of $\sigma~=~1^{\circ}43 \pm 0^{\circ}17_{\mbox{\scriptsize stat}}$.
The difference in TS between this model and the initial one is 45.
Given the high number of degrees of freedom between this model and the initial model, we can use the Akaike Information Criterion (AIC;~\citealt{AIC}) given by AIC = $2k - 2\mbox{ln}\mathcal{L}$ with $k$ being the number of free parameters and $\mathcal{L}$ being the maximum value of the likelihood function. This penalizes the model with the largest number of free parameters, so that the model with the fewest parameters will be favored, unless the extra parameters actually provide a substantially better fit. The best model is the one that have a lower AIC value. In this case, the four-component model is clearly preferred to the initial model, with $\Delta$AIC = 74. 
An excess of $\sim$3$\sigma$ significance remains at the top of the region of interest, visible on map (c) of Figure \ref{summary_b}. Adding a new component at its location improves the fit only by a $\Delta$TS of 10, which is not significant when considering the addition of another source with 4 degrees of freedom. The $\Delta$AIC is 12. Since there are no compelling counterparts to this excess at other wavelengths, the remaining excess may be the result of additional complexities that are not contained in our model, including spatial morphology asymmetries and more complex spectral shapes, or it may simply be due to fluctuations. However, even though it gives a similar likelihood value, using an asymmetric Gaussian shows a clear 5$\sigma$ signal in the residual map at the location of J1928. Moreover, neither a power law with exponential cutoff nor a log-parabola significantly improve the fit.

\newpage

\newpage

\begin{table}[ht!]
 \caption{ Input values and fitted values (subscripts $i$ and $f$ respectively) for each component of the best model representing the region of interest. Each value is followed by the statistical uncertainty and the systematic uncertainty. The fit is performed in two steps. The initial model has two components representing J1928 and J1930. A point-like component and an extended Gaussian component are added at the location of significant excess in the residual map. The flux normalization is given at 10~TeV in units of 10$^{-15}$~TeV$^{-1}$~cm$^{-2}$~s$^{-1}$. The spectral energy distributions are plotted in Figure~\ref{all_spec-PL}. The position of the pulsar PSR~J1930+1852 is (292\textdegree63, 18\textdegree87).
 }
 \centering
 \begin{tabular}{|c|c|c|c|c|c|}
                            & J1930                                 & J1932                                 & J1928                                 & J1928-EXT               \\ 
  \cline{1-5}
  Hypothesis                 & Point-like                            & Point-like                           & Extended                              & Extended                   \\
  \cline{1-5}
  pos${_i}$                  & PSR J1930+1852                        & (293.07, 19.40)                      & (292.08,17.79)                        & (292.20,18.18)             \\
  pos${_f}$ (ra \textdegree) & 292.53 {\tiny $\pm$ 0.05 $\pm$ 0.004} & 292.99 {\tiny $\pm$ 0.05 $\pm$ 0.002} & 292.15 {\tiny $\pm$ 0.04 $\pm$ 0.001} & 292.05 {\tiny $\pm$ 0.15 $\pm$ 0.05}  \\
  \qquad (dec \textdegree)   & 18.84 {\tiny $\pm$ 0.05 $\pm$ 0.001}  & 19.36 {\tiny $\pm$ 0.04 $\pm$ 0.001}  & 17.90 {\tiny $\pm$ 0.04 $\pm$ 0.001}  & 18.10 {\tiny $\pm$ 0.17 $\pm$ 0.05}   \\
  \cline{1-5}
  size${_i}$ (\textdegree)   &   -                                  & -                                     & 0.10                                  & 0.9                        \\
  size${_f}$  (\textdegree)  &   -                                  & -                                     & 0.18 {\tiny $\pm$ 0.04 $\pm$ 0.003}   & 1.43 {\tiny $\pm$ 0.17 $\pm$ 0.05}     \\
  \cline{1-5}
  index${_i}$                &  $-2.5$                              & $-2.5$                                 & $-2.5$                    & $-2.5$                       \\
  index${_f}$                &  $-2.93$ {\tiny $\pm$ 0.20 $\pm$ 0.01} & $-2.46$ {\tiny $\pm$ 0.24 $\pm$ 0.01} & $-2.09$ {\tiny $\pm$ 0.16 $\pm$ 0.04} & $-2.60$ {\tiny $\pm$ 0.08 $\pm$ 0.01}   \\
  \cline{1-5}
  flux$_{i}$                 & 10.0                      & 10.0                        & 10.0                      &  10.0                      \\
  \multirow{2}*{flux$_{f}$}  & \multirow{2}*{2.46 $_{-0.47}^{+0.58}$ {\tiny $\pm$ 0.72} } & \multirow{2}*{1.95 $_{-0.49}^{+0.62}$ {\tiny $\pm$ 0.50}} & \multirow{2}*{4.23 $_{-1.10}^{+1.49}$ {\tiny $\pm$ 1.30}} & \multirow{2}*{40.34 $_{-4.11}^{+4.47}$ {\tiny $\pm$ 1.93}} \\
   & & & & \\
  \cline{1-5}
  Energy range (TeV) & 1 -- 118 & 1 -- 43 & 1 -- 178 & 1 -- 10 \\
  \cline{1-5}

\end{tabular}

  \label{model_parameters}
\end{table}

\newpage

%-------------------------------------------------------------%

\subsection{Energy spectrum of 3HWC J1930+188} 

\begin{figure}[ht!]
    \centering
    \includegraphics[width=0.5\linewidth]{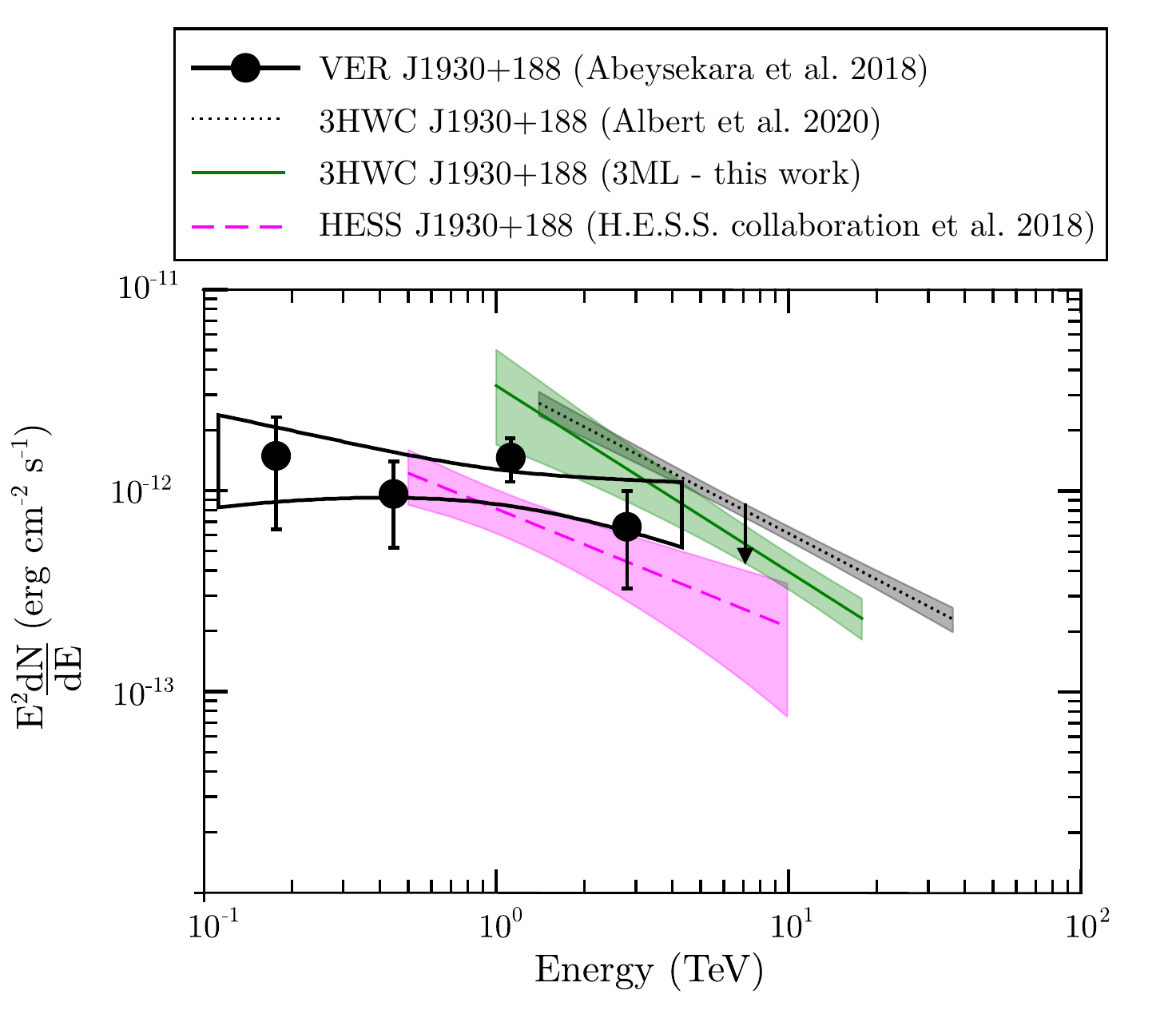}
    \caption{ Energy spectrum of 3HWC~J1930+188. The green spectrum is the result from the fit of a point-like component for 3HWC~J1930+188, as part of a model of the region assuming two point sources and two extended sources. The gray spectrum uses the same amount of data, the analysis bins 1 to 9, and a single point-like source hypothesis~\citep{3HWC_catalog}. The H.E.S.S. spectrum is depicted in magenta and is taken from the HGPS~\citep{HGPS}. The shaded areas represent the 1$\sigma$ statistical uncertainty. The black dots are derived by the  VERITAS collaboration~\citep{Veritas_Fermi_2HWCsources}. All the spectral parameters for the spectra plotted here are gathered in Table~\ref{J1930_spec_parameters}.
    }
    \label{hawc-veritas-hess_spectrum}
\end{figure}

The spectrum of 3HWC J1930+188 resulting from the fit of the four-component model described in the previous section is shown in green in Figure~\ref{hawc-veritas-hess_spectrum}. The spectrum is slightly softer than the one previously published by the HAWC collaboration, using the same amount of data, analysis bins 1 to 9, and a single point-like source hypothesis, shown in gray~\citep{3HWC_catalog}, while in the present analysis it is part of a more complex model. The high number of free parameters being fitted together is responsible for the larger uncertainties. At a few TeV, the spectrum from the fit presented here is in better agreement with the VERITAS spectrum~\citep{Veritas_Fermi_2HWCsources}, represented by the black dots, although the error bars are wider. The H.E.S.S. spectrum~\citep{HGPS} is also shown in magenta. The spectral parameters derived in the different works cited here are gathered in Table~\ref{J1930_spec_parameters}.

\begin{table}[ht!]
 \caption{ Spectral parameters and their statistical uncertainties for the spectral energy distributions of 3HWC J1930+188 plotted in Figure~\ref{hawc-veritas-hess_spectrum}.
 }
 \centering
 \begin{tabular}{|l|c|c|c|c|}
 \hline 
 \multirow{2}*{Experiment (reference)} & Reference Energy & Flux at E$_0$                              & \multirow{2}*{Index} & Integrated Flux $>$ 1~TeV \\
                                       & E$_0$ (TeV)      & ($10^{−15}$ TeV$^{−1}$ cm$^{−2}$ s$^{−1}$) &                      & (10$^{-12}$ cm$^{-2}$ s$^{-1}$)    \\
 \hline
 HAWC (this work) & 10 & 2.46 $_{-0.47}^{+0.58}$ & $-2.93\pm0.20$ &  $1.08 \pm 0.46$\\
 HAWC \citep{3HWC_catalog} & 7 & $10.2\pm0.8$ & $-2.76\pm0.07$ & $1.25 \pm 0.16$ \\
 VERITAS \citep{Veritas_Fermi_2HWCsources} & 1 &  $660\pm130$  & $-2.18\pm0.20$ & $0.57 \pm 0.14$\\
 H.E.S.S. \citep{HGPS} & 1 & $506\pm124$ & $-2.59\pm0.26$ & $0.32 \pm 0.09$\\
 \hline
 \end{tabular}

  \label{J1930_spec_parameters}
\end{table}

%-------------------------------------------------------------%

\subsection{Characteristics of the new source HAWC J1932+192} 
In this section, we discuss whether the $\gamma$-ray emission of the new TeV source candidate HAWC~J1932+192, potentially associated with the pulsar PSR~J1932+1916, could come from the acceleration of particles in its PWN. All characteristics of this system have been previously gathered in section~\ref{Multi-wavelength observations of the region}, as well as in the appendix~\ref{Multi-wavelength information}, Table~\ref{J1932_caracteristics}. The spectrum derived from 3ML under the point-like hypothesis is plotted in Figure~\ref{all_spec-PL}.

From the best fit, the differential flux at 10~TeV was found to be  $(1.95 _{-0.49}^{+0.62})_{\mbox{\scriptsize stat}}\times10^{-15}$~TeV$^{-1}$~cm$^{-2}$~s$^{-1}$. With a spectral index equal to $-2.46\pm0.24$, the integrated energy flux between 1~TeV and 43~TeV is F$_{\gamma > 1 \mbox{\tiny TeV}}~=~(1.61\pm0.58)_{\mbox{\scriptsize stat}}\times10^{-12}$~erg~cm$^{-2}$~s$^{-1}$. 
The $\gamma$-ray luminosity is given by:
\begin{equation}
 L_{\gamma} = 4\pi D^2 F_{\gamma>1\mbox{\tiny TeV}}.
 \label{luminosity}
\end{equation}
Under the assumption that the distance is either 3.5~kpc~\citep{G54.4_CO} or 6.6~kpc~\citep{G54.4}, we can calculate the $\gamma$-ray luminosity $L_{\gamma}$ and since the pulsar's rotational energy is $\dot{E} = 4\times10^{35}$ erg s$^{-1}$, we can calculate the energy that the pulsar has to spend to produce it. Using the first distance estimation, $\sim$0.6\% of the pulsar energy is needed to accelerate the electrons and positrons that produce the $\gamma$ rays via IC scattering on ambient photons. In the case of the larger distance, this percentage goes up to $\sim$2\%. For comparison, ~\cite{Geminga_fermi} found that about 1\% of the spin-down energy of the Geminga pulsar has to be converted into e$^{\pm}$ to be consistent with the $\gamma$-ray data from the Fermi-LAT and from HAWC.  
This means that the PWN could in principle produce the observed $\gamma$-ray emission. Table~\ref{J1932_parameters} gathers the parameters calculated above.

\begin{table}[ht!]
 \caption{Summary of the properties of the new source HAWC~J1932+192 .}

 \centering
 \begin{tabular}{|c|c|c|}
  \hline 
  Morphology Hypothesis                                                    & \multicolumn{2}{c|}{Point-like} \\     
  \hline
  \multirow{2}*{Integrated energy flux F$_{\gamma > 1 \mbox{\tiny TeV}}$ (erg~cm$^{-2}$~s$^{-1}$)} & \multicolumn{2}{c|}{\multirow{2}*{($1.61\pm0.58)_{\mbox{\scriptsize stat}}\times10^{-12}$}} \\
                                                                          &\multicolumn{2}{c|}{}        \\
  \hline 
  Distance $D$ (kpc)                                                        & $\sim3.5$        & 6.6      \\
  \hline
  \multirow{2}*{$\gamma$-ray luminosity $L_{\gamma}$ (erg~s$^{-1}$)}       & \multirow{2}*{$\sim2.4\times10^{33}$}  &  \multirow{2}*{$\sim8.5\times10^{33}$}\\   
                                                                          &                               &  \\             
  \hline
  Fraction of the pulsar energy needed (\%)                                              & $\sim 0.6$               & $\sim 2$ \\
  \hline
\end{tabular}
  \label{J1932_parameters}
\end{table}

%-------------------------------------------------------------%
\subsection{Morphology and energy spectrum of 3HWC J1928+178} 

\begin{figure}[ht!]
    \centering
    \includegraphics[width=0.5\linewidth]{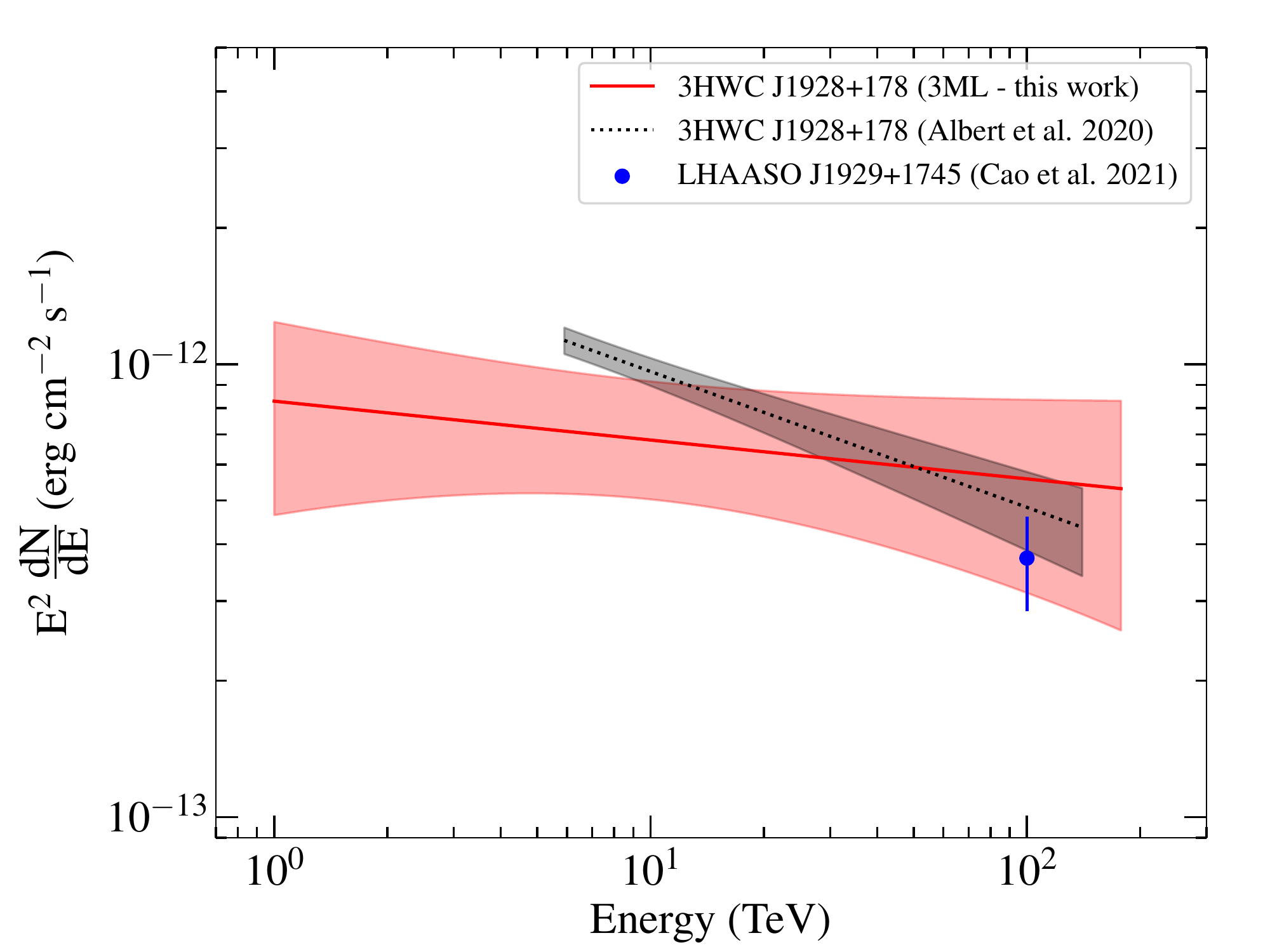}
    \caption{ Energy spectrum of 3HWC~J1928+178 from the 3ML fit, assuming a 2D Gaussian (red,) compared to that from~\citet{HAWC-HESS_GP_ApJ}, assuming a point-like source (gray). Both make use of the same data set, but different analysis bins. The shaded areas represent the 1$\sigma$ statistical uncertainties. The blue point is the flux point reported by LHAASO~\citep{LHAASO}. 
    }
    \label{J1928_spectrum}
\end{figure}

The best fit of the data gives a size of $\sigma=0^{\circ}18
\pm 0^{\circ}04_{\mbox{\scriptsize stat}}$ for 3HWC J1928+178, which represents 39\% containment, given our 2D Gaussian model. The corresponding 68\% containment radius is 0\textdegree27.
The flux at 10~TeV is $(4.23_{-1.10}^{+1.49})_{\mbox{\scriptsize stat}}\times10^{-15}$~TeV$^{-1}$~cm$^{-2}$~s$^{-1}$ and the spectral index is $-2.09\pm0.16$, as reported in Table \ref{model_parameters}. 
The spectrum is plotted in red in Figure~\ref{J1928_spectrum} together with the one previously published by the HAWC collaboration, in gray, using the same amount of data and analysis bins 1 to 9, but for a single point-like source hypothesis~\citep{3HWC_catalog}. As previously mentioned, the high number of free parameters being fitted together is responsible for the larger uncertainties. Both HAWC spectra are compatible with the flux point from LHAASO at 100 TeV~\citep{LHAASO} within the uncertainties. The origin of the observed TeV $\gamma$-ray emission of 3HWC~J1928+178 is discussed in the next section. A classical PWN scenario is considered, as well as a possible association with a molecular cloud. 
Note that the presence of the large extended component J1928-EXT of angular size $\sigma~=~1.43^{\circ} \pm 0.17^{\circ}_{\mbox{\scriptsize stat}}$ may account for a large-scale galactic diffuse emission component that is absent from the model, or may indicate the mismodeling of 3HWC J1928+178. In particular, J1928 and J1928-EXT may be part of the same object, if we consider a Geminga-like diffusion model. This hypothesis was considered in~\cite{3HWCJ1928_ICRC} and will not be treated here.\\

%%%%%%%%%%%%%%%%%%%%%%%%%%%%%%%%%%%%%%%%%%%%%%%%%%%%%%%%%%%%%%%%
\section{Origin of the $\gamma$-ray emission of 3HWC J1928+178} 
\label{Origin}

\subsection{IC scattering of the electrons from the PWN}
\paragraph{Gamma-ray emission} 
From the fitted size of 3HWC J1928+178, $\sigma = 0.18$\textdegree, the diameter $d$ and volume $V$ can be calculated assuming a spherical geometry. All the properties derived hereafter are summarized in Table \ref{PWN_parameters}.
With the pulsar being located at a distance of $D = 4.3$~kpc, 39\% and 68\% of the emission are contained in regions of sizes $d \simeq$ 27~pc and 41~pc, respectively.
The integrated energy flux between 1~TeV and 178~TeV is $F_{\gamma>1\mbox{\tiny TeV}}~=~(3.45\pm1.22)_{\mbox{\scriptsize stat}}\times10^{-12}$~erg~cm$^{-2}$~s$^{-1}$ . 
The $\gamma$-ray luminosity, given by equation~\ref{luminosity}, is $L_{\gamma} = 7.7\times10^{33}$ erg~s$^{-1}$. 

The emission observed in PWNe at TeV energies is dominated by radiation processes involving electrons scattering on ambient photons: IC scattering. In the Thomson regime, the $\gamma$-ray spectral energy distribution due to electrons with energy $E_e$ peaks at
\begin{equation}
 E_{\gamma} \simeq 33E_e^2k_BT \quad \mbox{TeV,}
  \label{E_e_vs_E_g_thomson}
\end{equation}
where $E_{\gamma}$ and $E_e$ are in TeV, $k_B$ is the Boltzmann constant, T is the temperature of the photon field, and $k_BT$ is in eV~\citep{TeV_astronomy}.
Hence, $ E_e \simeq  11\sqrt{E_{\gamma}} ~\mbox{TeV}$
and a 1~TeV $\gamma$-ray photon is produced via IC scattering of an electron of energy $\sim$10~TeV on cosmic microwave background (CMB) photons. 
The electron cooling time for IC scattering in the Thomson regime is given by
\begin{equation}
 \tau_{\mbox{\tiny IC}} = \frac{E_e}{dE_e/dt} \simeq 3.1\times10^5\frac{1}{U_{\mbox{\tiny rad}}}\frac{1}{E_e} \quad \mbox{yr,}
 \label{IC_cooling_time}
\end{equation}
where $E_e$ is in TeV and $U_{\mbox{\tiny rad}}$ is the radiation energy density in eV~cm$^{-3}$~\citep{TeV_astronomy}. For electrons of energy $E_e$~=~$10$~TeV scattering on CMB photons, $k_BT~=~2.35\times10^{-4}$~eV and $U_{\mbox{\tiny rad}}~=~0.26$~eV~cm$^{-3}$, so the electron cooling time is $\tau_{\mbox{\tiny CMB}} \simeq 120$~kyr. 
For far-IR (FIR) photons,  $k_BT~=~3\times10^{-4}$~eV and $U_{\mbox{\tiny rad}}~=~0.3$~eV~cm$^{-3}$, so $\tau_{\mbox{\tiny FIR}} \simeq 100$~kyr. 
The total energy is the product of the $\gamma$-ray luminosity and the cooling time 
$W = \tau_{\mbox{\tiny IC}}~L_{\gamma}$, equal to $W_{\mbox{\tiny CMB}} \simeq 2.9\times10^{46}$ erg, using $\tau_{\mbox{\tiny IC}} = \tau_{\mbox{\tiny CMB}}$. Finally, dividing by the volume, the energy density is simply $\epsilon_{\mbox{\tiny W}} = W/V$. 
Assuming a spherical geometry and a diameter of 41~pc, the energy density is $\epsilon_{\mbox{\tiny IC}} \simeq 0.04$ eV~cm$^{-3}$. This is much smaller than the energy density of the interstellar medium (ISM) $\epsilon_{\mbox{\tiny ISM}} \simeq 1$~eV~cm$^{-3}$. Given the age of the pulsar of 82~kyr, this is consistent with an old PWN, where the electrons have started to cool and diffuse away from their source. Note that for electrons with $E_e>300$~TeV, the Klein Nishina regime starts. 
Adapting equations~\ref{E_e_vs_E_g_thomson} and \ref{IC_cooling_time} to the Klein Nishina regime for CMB photons only gives 300 TeV electrons producing 230~TeV photons, with the cooling time becoming $\tau_{\mbox{\tiny CMB}} \simeq 30$~kyr. However, the total energy and the energy density are of the same order of magnitude as what was calculated in the Thomson regime. 

\begin{table}[ht!]
 \caption{Summary of the properties of the fitted source J1928 in the hypothesis where IC scattering on CMB photons is the dominant radiation process.}
 \centering
 \begin{tabular}{|c|c|}
      
  \hline 
  \multirow{2}*{Angular size $\theta$ (\textdegree)}                       & 0.18 (39\%)          \\   
                                                                           & 0.27 (68\%)                   \\     
  \hline
  Distance $D$ (kpc)                                                        & 4.3         \\
  \hline
  Size (68\%) $d$ (pc)                                                     & $\sim41$           \\
  \hline
  \multirow{2}*{Volume $V$ (pc$^{3}$)}                                     & \multirow{2}*{$\sim3.7\times10^{4}$}        \\   
                                                                           &                                 \\                       
  \hline 
  \multirow{2}*{Integrated energy flux F$_{\gamma > 1 \mbox{\tiny TeV}}$ (erg~cm$^{-2}$~s$^{-1}$)} & \multirow{2}*{($3.45\pm1.22)\times10^{-12}$} \\
    &   \\
  \hline
  \multirow{2}*{$\gamma$-ray luminosity $L_{\gamma}$ (erg~s$^{-1}$)}       & \multirow{2}*{$\sim7.7\times10^{33}$}        \\   
                                                                           &                                 \\                       
  \hline 
  \multirow{2}*{Total energy $W_{\mbox{\tiny IC}}$ (erg)}                  & \multirow{2}*{$\sim2.9\times10^{46}$}       \\     
                                                                           &                                  \\
  \hline
  \multirow{2}*{Energy density $\epsilon_{\mbox{\tiny IC}}$ (eV~cm$^{-3}$)} &  \multirow{2}*{$\sim$0.04}              \\      
                                                                            &                                    \\
  \hline
\end{tabular}
  \label{PWN_parameters}
\end{table}

\paragraph{Parent particle population} 
The parent population of the electrons responsible for the observed $\gamma$-ray emission can be obtained using the \textit{naima}\footnote{The documentation and code for naima are available at: https://naima.readthedocs.io/en/latest and https://github.com/zblz/naima} python package~\citep{naima}. This provides models for nonthermal radiative emission from homogeneous distributions of relativistic particles. The contributions of nonthermal radiative processes, IC scattering in this case, can be computed given a shape for the particle energy distribution, and the model can be used to fit the observed nonthermal spectra through a Markov Chain Monte Carlo procedure. In the present case, the emission is assumed to be produced by electrons upscattering CMB photons, with a temperature $T = 2.72$~K and an energy density of 0.26~eV~cm$^{-3}$, and FIR photons, with a temperature $T = 20$~K and an energy density of 0.3~eV~cm$^{-3}$. Since the $\gamma$-ray spectrum of 3HWC~J1928+178 has been represented by a power law, the population of the electrons is also chosen to follow a simple power law. The fit is performed using this model for the electrons and the $\gamma$-ray spectrum from the HAWC observations from the best fit with 3ML.
The best fit for the energy distribution of electrons has a differential energy at 70 TeV $F_{70\mbox{\tiny TeV}}~=~(1.91~\pm~0.2)\times10^{41}$~erg$^{-1}$ and an index of $-2.55~\pm~0.1$. The total energy of the electrons above 1~TeV is $W_e~=~4.6^{+2.2}_{-1.2}\times10^{46}$~erg. Given that the spin-down of the pulsar is $\dot{E}~=~1.6\times10^{36}$~erg~s$^{-1}$, assuming that it is constant over the life of the pulsar, which is 82~kyr, gives a lower limit for the total energy released by the pulsar of $4.1\times10^{48}$~erg. Hence, an upper limit of $\sim$1\% can be set on the amount of energy that the pulsar could have transferred to the electrons above 1 TeV. This is again compatible with previous estimations for the Geminga pulsar~\citep{Geminga_fermi}.

\subsection{Association with a molecular cloud} 
\paragraph{Hypotheses for this association} 
Most of the interstellar gas in our Galaxy is molecular hydrogen H$_{2}$, contained in GMCs. These massive clouds of gas and dust have a typical size that ranges from 50 to 200 pc and a mass ranging between $10^4$ and $10^6$ solar masses. They are the sites of star formation. In addition, they are the source of most of the diffuse galactic $\gamma$-ray emission~\citep{EGRET_diffuse_emission}. The dominant processes by which cosmic rays interact with the ISM and produce $\gamma$ rays, are high-energy electron bremsstrahlung, IC interactions with low-energy photons and nucleon-nucleon interactions. For the latter, in particular, molecular clouds are favorable environments. Hence, it is interesting to compare the galactic gas distribution, and the $\gamma$-ray emission detected by HAWC, in order to assess whether the components of the molecular cloud,  mainly hydrogen, could be a target for relativistic protons, producing observed $\gamma$~rays via pion decay~\citep{HAWC_GMC}.

\paragraph{CO as a tracer for molecular clouds}
H$_{2}$ is not easily observable, because this molecule has no electric dipole moment. 
For this reason, it does not emit radiation from neither vibrational nor rotational transitions. However, CO emits radiation through a rotational transition ($J=1\rightarrow0$) when excited by collisions with hydrogen molecules. Hence, CO emission is used to trace H$_{2}$ molecular clouds. The abundance of CO is typically about $7.2\times10^{-5}$ for one hydrogen molecule. Two isotopes are mainly used: $^{12}$CO and $^{13}$CO. The main difference is that $^{12}$CO is optically thick, while $^{13}$CO is optically thin, the first one being on average $\sim$60 times more abundant than the second one~\citep{12CO_13CO_ratio}. 
$^{13}$CO is both a good quantitative and qualitative tracer of molecular gas, being related to the column density of H$_2$. It can probe deep in the cloud without saturating, and it provides more accurate velocity and kinematic distances because of its narrower line. Therefore, $^{13}$CO is more suited to deriving the column density of the cloud, under the hypothesis of local thermodynamic equilibrium.
The emission line corresponding to the CO de-excitation gives the mean velocity of the CO molecules in the cloud, and the width of this line gives the velocity dispersion associated with the cloud. Under the virial equilibrium hypothesis, and assuming uniform density within the cloud, the width of the line scales linearly with the size of the cloud.

\begin{figure}[ht!]
    \centering
    \includegraphics[width=1\linewidth]{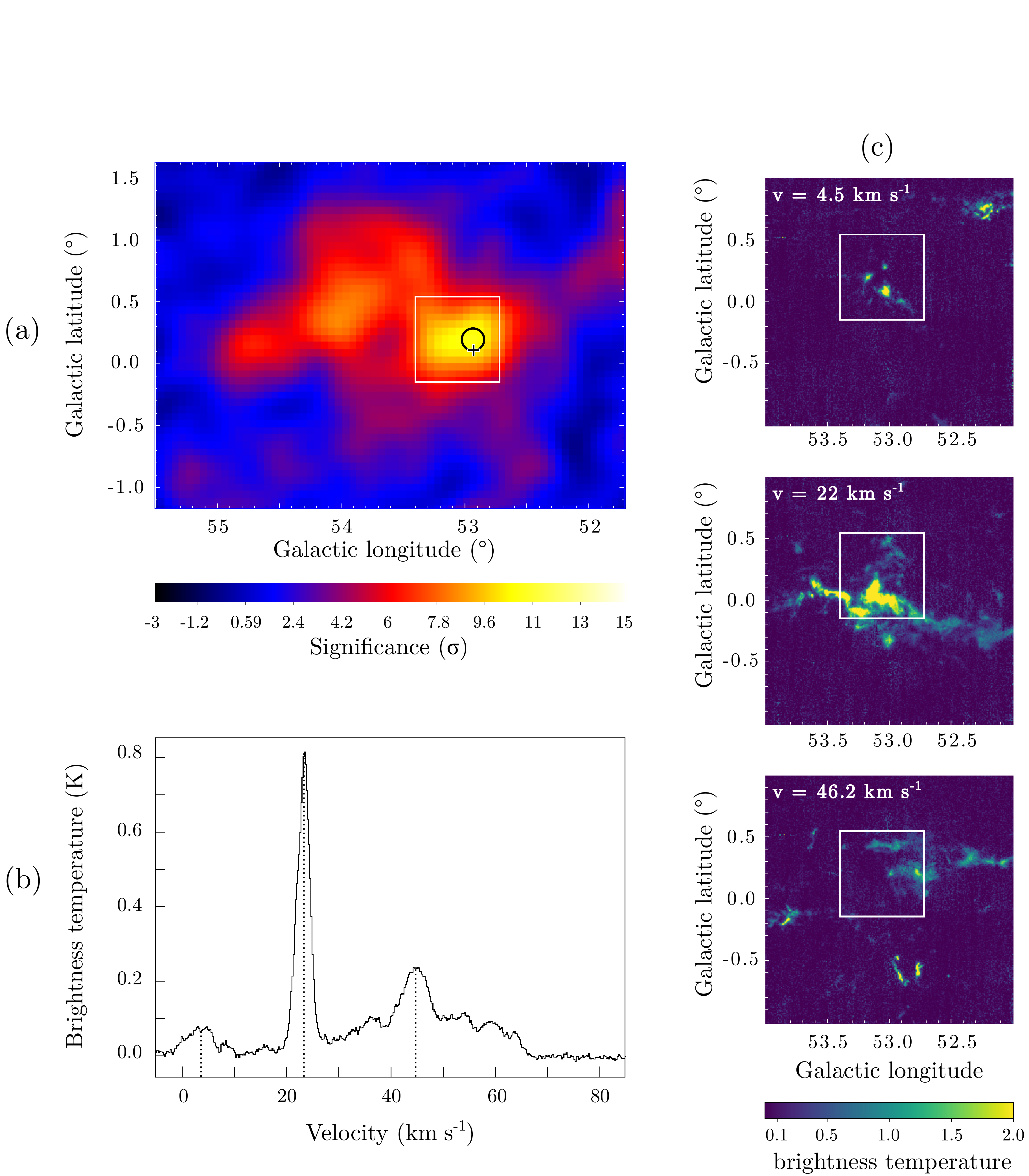}
    \caption{The HAWC significance map is shown in panel (a). The black circle shows the location and 1$\sigma$ uncertainty of the HAWC source~\citep{3HWC_catalog}. The black cross is the location of the pulsar PSR J1928+1746. The white box is the region with $>5\sigma$ $\gamma$-ray emission of 3HWC~J1928+178, with the velocity dispersion being averaged and plotted in panel (b) as a function of the brightness temperature. For the three peaks at $\sim$4.5~km~s$^{-1}$, $\sim$22~km~s$^{-1}$ and $\sim$46~km~s$^{-1}$, the $^{13}$CO maps are shown in panel (c).}
    \label{HAWC_CO_velocity}
\end{figure}

The $^{13}$CO (rotation emission line $J=1\rightarrow0$ at 110~GHz) data from the Galactic Ring Survey (GRS\footnote{GRS data available at : https://www.bu.edu/galacticring/new\_data.html}; \citealt{GRS}) were obtained using the SEQUOIA multi pixel array on the Five College Radio Astronomy Observatory (FCRAO, \citet{FCRAO}) 14~m telescope located in New Salem, Massachusetts, between 1998 December and 2005 March. Three molecular clouds can be found at the location of the HAWC TeV emission. Figure \ref{HAWC_CO_velocity}(a) shows the HAWC significance map, where a region corresponding to the emission with significance $>5\sigma$ is defined. From this region, the velocity distribution is extracted as a function of the brightness temperature averaged over this region, visible in Figure~\ref{HAWC_CO_velocity}(b).  
Three maxima can be highlighted at $\sim$4.5~km~s$^{-1}$, $\sim$22~km~s$^{-1}$ and $\sim$46~km~s$^{-1}$. The $^{13}$CO maps corresponding to each velocity are also displayed in Figure \ref{HAWC_CO_velocity}(c). The most intense one, at $\sim$22~km~s$^{-1}$ is further studied in the next paragraph.

\begin{figure}[ht!]
    \centering
    \includegraphics[width=1\linewidth]{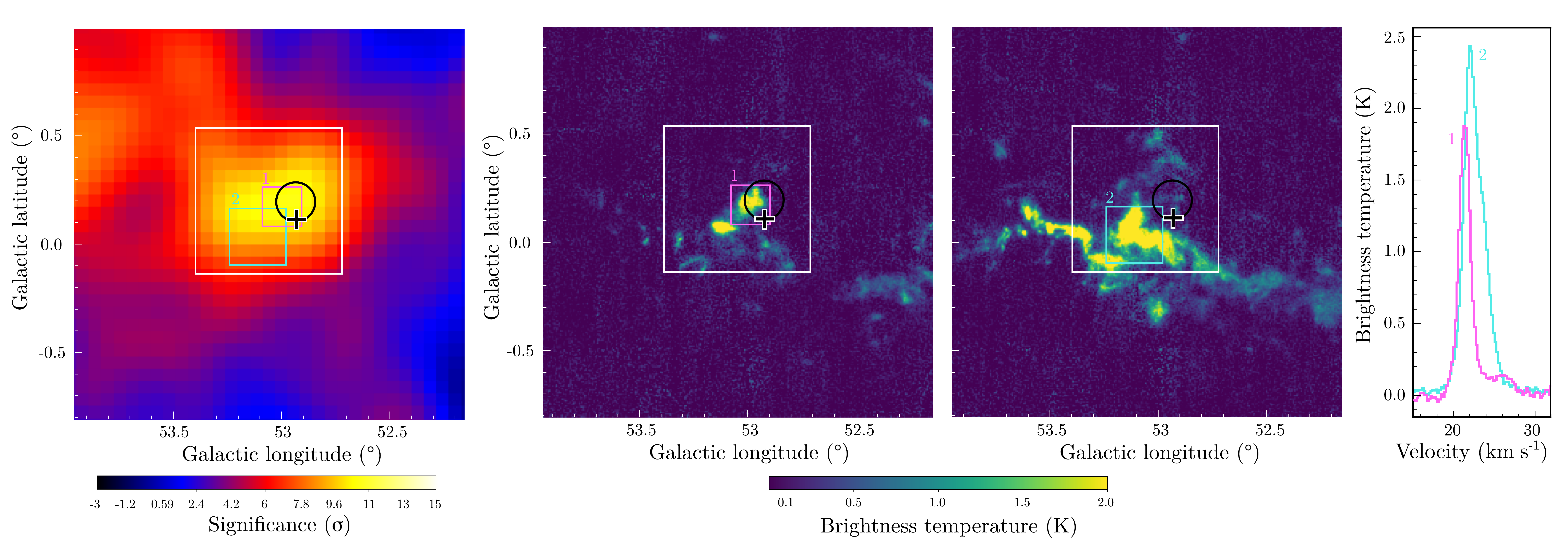}
    \caption{Molecular clouds at 22 km~s$^{-1}$ located within the 5$\sigma$ $\gamma$-ray emission of 3HWC~J1928+178 (white box). The black circle shows the location and 1$\sigma$ uncertainty of the HAWC source~\citep{3HWC_catalog}. The black cross is the location of the pulsar. The  magenta and cyan boxes correspond to the two clumps considered here. The two velocity maps corresponding to the two boxes are shown in the center, with the velocity dispersion on the right-hand side.}
    \label{HAWC_CO_velocity_22kms}
\end{figure}

\paragraph{Detailed study of the brightest cloud}
The cloud at $\sim$22~km~s$^{-1}$ has a very complicated and elongated shape. The study is restricted to the portion of the cloud within the $>5\sigma$ $\gamma$-ray emission of 3HWC~J1928+178, represented by the white box in Figures \ref{HAWC_CO_velocity} and \ref{HAWC_CO_velocity_22kms}. In this region, the cloud is decomposed into two parts, which could be interpreted as two clumps of the cloud. They are represented by the two smaller magenta and cyan boxes that are labeled ``1" and ``2" in Figure \ref{HAWC_CO_velocity_22kms}. The $^{13}$CO maps for the peak velocity and the velocity distribution are also shown on the right-hand side of the same figure. Some basic properties can now be derived, such as the column density, the mass, and the volume of these clumps, to estimate the total cosmic-ray energy and the energy density that would be needed to explain the observed $\gamma$-ray emission. 

To do so, the clumps are assumed to have a spherical shape. The most probable distance\footnote{The distance is derived using http://bessel.vlbi-astrometry.org/bayesian with a prior P$_{far}$ = 0.1} for this cloud is $D$ = 4~kpc~\citep{cloud_distance}, which would be compatible with the distance of the pulsar. Their diameter $d$ and volume $V$ can be calculated from their angular size $\theta$. Clumps 1 and 2 have diameters of $\sim$12 and $\sim$18~pc, respectively. They are smaller than the source representing J1928, which was found to contain 68\% of the emission within $\sim$41~pc.  
For each clump, the $^{13}$CO column density $N(^{13}$CO) is determined using the brightness temperature $T_{\mbox{\scriptsize mb}}$, in K, and the FWHM of the velocity distribution peak $\Delta v$, in km~s$^{-1}$, as explained in \citet{Simon_molecular_cloud}:
\begin{equation}
 N(^{13}\mbox{CO}) = 8.75\times10^{14} T_{\mbox{\scriptsize mb}} \Delta v.
\end{equation}
The clump mass $M$, in the unit of solar masses, is given by
\begin{equation}
 M = 3.05\times10^{-25} N(^{13}\mbox{CO})~\theta_x \theta_y D^2 \quad M_{\odot},
\end{equation}
where $\theta_x$ and $\theta_y$ are the half-axes of the clump in arcseconds and $D$ is the distance to the cloud in~pc, which is here assumed to be 4~kpc. With the mass and the volume, the particle density in the cloud, which is a potential target for cosmic rays, can be calculated using
\begin{equation}
 n = \frac{M}{\mu m_{\mbox{\tiny H}}V},
\end{equation}
where $\mu m_{\mbox{\tiny H}}$ is the mean mass of an atom in the ISM, with $\mu\simeq1.4$ and $m_{\mbox{\tiny H}}$ being the mass of an hydrogen atom.
Moreover, using the best-fit value for the flux found with 3ML, the luminosity $L_{\gamma}$ above~1~TeV was calculated in the previous section using equation \ref{luminosity} as $L_{\gamma} \simeq 7.7\times10^{33}$~erg~s$^{-1}$.
Considering that, at TeV energies, the spectral energy distribution of the secondary $\gamma$ rays peaks at about one-tenth of the energy of the primary proton and does not vary significantly with energy~\citep{TeV_astronomy}, a 1~TeV photon can be produced by a 10~TeV proton. 
The total energy of the cosmic rays above~10~TeV in the cloud is $W_{\mbox{\tiny p}} = \tau_{\mbox{\tiny p}} L_{\gamma}$, where $\tau_{\mbox{\tiny p}}$ is now the characteristic cooling time for relativistic protons. It is derived using the proton-proton interaction cross section $\sigma_{\mbox{\tiny pp}}$, the speed of light $c$ and the density $n$:
\begin{equation}
 \tau_{\mbox{\tiny p}} = \frac{1}{f\sigma_{\mbox{\tiny pp}}cn}.
\end{equation}
In this relation, $f$ stands for the fact that a proton loses about half of its energy per interaction, with only a third of them producing $\pi^0$. Hence, using typical values for the inelastic cross section $\sigma_{\mbox{\tiny pp}} \simeq 35$~mb for VHE protons \citep{TeV_astronomy} and $f = 1/6$, it results in a lifetime $\tau_{\mbox{\tiny p}}~\simeq~1.8\times10^8~n^{-1}$~yr. 
Finally, the energy density is simply the ratio of the total energy and the volume:
$ \epsilon_{\mbox{\tiny p}} = W_{\mbox{\tiny p}}/V$.
For the cloud considered here, the total energy of the cosmic rays above~10~TeV is $W_{\mbox{\tiny p}} = 7.9\times10^{47}$~erg and the energy density is $\epsilon_{\mbox{\tiny p}} \simeq 4.4$~eV~cm$^{-3}$.
The parameters calculated for each clump and for the total cloud are gathered in Table~\ref{CO_clumps_parameters}. 

The farthest edge of clump 2 is 22~pc away from the pulsar. Considering a sphere of radius 22 pc centered on the pulsar, containing both clumps, its volume is 15 times the sum of the volumes of both clumps together. Since the total energy in the cloud is $W_{\mbox{\tiny p}} = 7.9\times10^{47}$~erg, the energy in the sphere centered on the pulsar should be $W_R = 1.2\times10^{49} \mbox{erg}$.
The pulsar releases most of its energy at the beginning of its lifetime and steadily decreases its spin-down power afterward, as described by the following equation:
 \begin{equation}
  \dot{E} = \dot{E_0}\Big( 1 + \frac{t}{\tau_{\tiny 0}} \Big)^{-\frac{n+1}{n-1}}, 
  \label{spin_down_luminosity}
 \end{equation}
where $\dot{E}_0$ is the initial spin-down luminosity and $n$ the braking index. It has been argued that up to 20\% of a pulsar's energy could accelerate ions~\citep{20percent_limit}.
Assuming as a reasonable value that 10\% of the pulsar's energy has been used to accelerate protons, this means that the pulsar must have released $10 \times W_R = 1.2\times10^{50}$~erg. Considering that this is equal to the difference in rotational energy between now and when the pulsar was born gives:
$ \Delta \mbox{E} = 1.2\times10^{50} =  I \times (\Omega^2_0 - \Omega^2) /2$, 
with $I\simeq 1 \times 10^{45}~\mbox{g cm}^2$ for a typical pulsar.
The pulsar considered here has a period of $P\simeq70$~ms. This gives a birth period of $P_0 = 2\pi/\Omega_0\simeq10$~ms. 
The maximum total energy that a pulsar with a birth period of 1~ms can release during its life is $\mbox{E}_{\mbox{\tiny ROT}} = 1\times10^{53}$~erg, for a pulsar with a typical mass of 1.4~$M_{\odot}$ and a typical radius of 10~km~\citep{pulsar_total_energy}. Our result is consistent with this upper limit. Moreover, integrating equation~\ref{spin_down_luminosity} from birth ($t=0$) until now ($t=T$), with the braking index $n=3$, 
and using the relation between the characteristic age of the pulsar $\tau_c$ and the age at birth 
$\tau_{\tiny 0} = \tau_c - \tau$
we can derive
\begin{equation}
 \tau_0 = \frac{\dot{E}~\tau_c^2}{\Delta E + \dot{E} \tau_c }.
\end{equation}
With $\Delta E = 1.2\times10^{50}$~erg, $\dot{E}~=~1.6\times10^{36}$~erg~s$^{-1}$ and $\tau_c = 82600$~yr, the age at birth is $\tau_0\simeq 2700$~yr. Hence, the pulsar's true age would be 79,800~yr. Finally, its spin-down power at birth would be $\dot{E}_0~=~1.4\times10^{39}$~erg~s$^{-1}$. As a comparison, this is the same order of magnitude as the Crab, which makes this value plausible.

\begin{table}[ht!]
 \centering
  \caption{Summary of the properties of the CO cloud}

 \begin{tabular}{c|c|c|}
                                                                                                 & Clump 1                           & Clump 2                          \\ 
  \hline 
  \multicolumn{1}{|c|}{Angular size $\theta$ (\textdegree)}                                     & 0.172                             & 0.252            \\
  \hline
  \multicolumn{1}{|c|}{Distance $D$ (pc)}                                                        & 4000               & 4000               \\
  \hline
  \multicolumn{1}{|c|}{Size $d$ (pc)}                                                           & 12.0                & 17.6            \\
  \hline
  \multicolumn{1}{|c|}{Volume $V$ (pc$^{3}$)}                                                    & 906            & 2851             \\
  \hline 
  \multicolumn{1}{|c|}{Average brightness temperature $T_{mb}$ (K)}                             & 1.875             & 2.44                \\
  \hline
  \multicolumn{1}{|c|}{FWHM of the velocity distribution peak $\Delta v$ (km~s$^{-1}$)}         & 1.5               & 2.56                          \\
  \hline  
  \multicolumn{1}{|c|}{\multirow{2}*{Column density $N(^{13}$CO) (cm$^{-3}$)}}                   & \multirow{2}*{$2.46\times10^{15}$}     & \multirow{2}*{$5.47\times10^{15}$}       \\
  \multicolumn{1}{|c|}{}                                                                         &                                   &                                      \\
  \hline
  \multicolumn{1}{|c|}{Mass $M$ ($M_{\odot}$)}                                                   & 1151             & 5482                    \\
  \hline                                                        
  \hline
                                                                    &\multicolumn{2}{c|}{Total cloud} \\
                                                                    
  \hline  
  \multicolumn{1}{|c|}{Mass $M$ ($M_{\odot}$)}                            &\multicolumn{2}{c|}{6633 }  \\      
  \hline                                                                    
  \multicolumn{1}{|c|}{Volume $V$ (pc$^{3}$)}                              &\multicolumn{2}{c|}{3757}  \\  
  \hline                                                                    
  \multicolumn{1}{|c|}{Density $n$ (particles~cm$^{-3}$)}                  &\multicolumn{2}{c|}{50}  \\  
  \hline                                                                    
  \multicolumn{1}{|c|}{\multirow{2}*{Total energy $W_{\mbox{\tiny p}}$ (erg)} }                            &\multicolumn{2}{c|}{\multirow{2}*{$7.9\times10^{47}$} } \\  
   \multicolumn{1}{|c|}{}                                                                 &\multicolumn{2}{c|}{\multirow{2}*{ } }     \\ 
  \hline                                                        
  \multicolumn{1}{|c|}{Energy density $\epsilon_{\mbox{\tiny p}}$ (eV~cm$^{-3}$)} &\multicolumn{2}{c|}{4.4}  \\  
  \hline 
  \hline                                                                    
\end{tabular}
  \label{CO_clumps_parameters}
\end{table}

%\newpage

\paragraph{Conclusions for the molecular cloud association}
From the study performed in this section, we can make conclusions regarding the different hypotheses:

\begin{itemize}
    \item The components of the molecular cloud,  mainly hydrogen, could be a target for relativistic protons from the pulsar PSR~J1928+1746 and its PWN, producing neutral pions during the interaction, which emit the observed $\gamma$~rays. The energy radiated by the pulsar was found to be compatible with the energy needed to produce the observed $\gamma$-ray luminosity via proton-proton interaction.
    
    \item 
    Adding up the two clumps gives a mass for the cloud of $\sim$6600~$M_{\odot}$, and a density of $\sim$50 particles per cm$^{3}$. However, the $\gamma$-ray emission around 1~TeV is dominated by IC scattering of electrons on the CMB for medium densities lower than $\sim$240 particles per cm$^{3}$~\citep{TeV_astronomy}. Hence, bremsstrahlung does not play a significant role here: the observed emission cannot be explained by the electrons and positrons from the PWN interacting with the atoms of the cloud via bremsstrahlung.
    
    \item 
    The total energy in cosmic rays $>10$~TeV in the cloud derived from the observed $\gamma$-ray luminosity is calculated as $7.9~\times~10^{47}$~erg, leading to an energy density of $\sim$4.4~eV~cm$^{-3}$. 
    This is three orders of magnitude higher than the energy density of the sea of galactic cosmic rays above 10~TeV, which is $\sim$1$\times10^{-3}$~eV~cm$^{-3}$~\citep{CR_sea_spectrum}. 
    Hence, these cosmic rays cannot explain the TeV emission observed by HAWC via interaction with the cloud. 

    \item The remaining hypothesis is that a local accelerator, for example an SNR, as yet undetected, is producing the detected VHE $\gamma$-ray emission. It is commonly assumed that an SNR releases $\sim$10$^{51}$~erg of kinetic energy in the ISM, and that 10\% of it - that is $\sim$10$^{50}$ erg - is used for cosmic-ray acceleration. Assuming a cosmic-ray spectrum between 1~GeV and 1~PeV, with an energy dependence E$^{-2}$, 33\% of the energy flux is found above 10~TeV, which makes $\sim$3.3$\times10^{49}$ erg. 
    The ratio of the volume around the SNR, which is uniformly filled with cosmic rays, and the volume of the cloud scales like the ratio of the energy contained in each volume:
    \begin{equation}
    \frac{D^3}{r^3} = \frac{3.3\times10^{49}}{7.7\times10^{47}} \simeq 40 \qquad \Rightarrow \qquad D = (40r^3)^{1/3},
    \end{equation}
    where $r=10$ pc is approximately the radius of the cloud and $D$ is the distance from the SNR to the farther edge of the clump. 
    Thus, an SNR located within a distance of $\sim$40~pc from the cloud would be able to account for the cosmic rays producing the observed TeV emission via interaction with the molecules of the cloud. 

\end{itemize}

%%%%%%%%%%%%%%%%%%%%%%%%%%%%%%%%%%%%%%%%%%%%%%%%%%%%%%%%%%%%%%%%
\section{Conclusion} 
\label{Conclusion}

This paper gives a detailed description and multiwavelength overview of this complex region of the Galactic plane at longitude 52\textdegree~ $<~\ell~<~55$\textdegree. Two sources, 3HWC~J1930+188 and 3HWC~J1928+178, has already been reported in the third HAWC catalog~\citep{3HWC_catalog} and one source, HAWC~J1932+192, is detected for the first time at TeV energies. A multicomponent fit was presented using 3ML.
\begin{itemize}
 \item 3HWC~J1930+188 is represented by a point-like source. Its spectrum is described by a simple power law with a flux at 10~TeV of $(2.46~(^{+0.58}_{-0.47})_{\mbox{\scriptsize stat}}\pm0.72_{\mbox{\scriptsize sys}})\times10^{-15}$~TeV$^{-1}$~cm$^{-2}$~ s$^{-1}$ and a spectral index of $-2.93\pm0.20_{\mbox{\scriptsize stat}}\pm0.01_{\mbox{\scriptsize sys}}$. The spectrum is in better agreement with the VERITAS spectrum than previous measurements~\citep{Veritas_Fermi_2HWCsources}. 
 
 \item HAWC~J1932+192 is represented by a point-like source. Its spectrum is described by a simple power law with a flux at 10 TeV of $(1.95~(^{+0.62}_{-0.49})_{\mbox{\scriptsize stat}}\pm0.50_{\mbox{\scriptsize sys}})\times10^{-15}$~TeV$^{-1}$~cm$^{-2}$~ s$^{-1}$ and a spectral index of $-2.46\pm0.24_{\mbox{\scriptsize stat}}\pm0.01_{\mbox{\scriptsize sys}}$.
 The $\gamma$-ray emission is energetically consistent with a PWN scenario. 
 
 \item 3HWC~J1928+178 is represented by an extended source of angular size $\sigma~=~0.18^{\circ}\pm 0.04^{\circ}_{\mbox{\scriptsize stat}}$\ (39\% containment). It has a hard spectrum with an index of $-2.09\pm0.16_{\mbox{\scriptsize stat}}\pm0.04_{\mbox{\scriptsize sys}}$, which would explain the fact that HAWC is more sensitive to detect this source than IACTs. Its flux at 10 TeV is $(4.23~(^{+1.49}_{-1.10})_{\mbox{\scriptsize stat}}\pm1.30_{\mbox{\scriptsize sys}})\times10^{-15}$~TeV$^{-1}$~cm$^{-2}$~ s$^{-1}$. We studied different hypotheses for the origin of the observed $\gamma$-ray emission and concluded that three scenarios would be possible:
    \begin{enumerate}
     \item e$^{\pm}$ from the PWN started to cool and diffuse away from it, producing $\gamma$ rays via IC scattering on ambient photons;
     \item cosmic-ray protons produced by the pulsar interacted with a nearby molecular cloud and produced $\gamma$ rays via proton-proton interaction; and
     \item there is another unknown accelerator, such as a nearby SNR, located within $\sim$ 40~pc. However, no hint of such a SNR has been observed at any wavelength.
    \end{enumerate}
\end{itemize}
    
Regarding 3HWC~J1928+178, for now, the first scenario may still be considered the most probable one. Due to the age of the pulsar, the lack of X-ray emission, the extended emission observed by HAWC, and the low energy density compared to the ISM, 3HWC~J1928+178 is a candidate for the TeV halo family~\citep{3HWCJ1928_ICRC}. It may also be in a transitional phase between a classical PWN and a TeV halo, and may help us to understand the late evolution stage of a PWN. The second and third scenarios cannot be ruled out, and more complex morphological and spectral analysis will be needed to help distinguish between them. The possibility that the $\gamma$-ray emission comes from protons produced by the pulsar interacting with a molecular cloud makes it a particularly interesting case to be followed up. However, this hypothesis relies on the estimated distance of this molecular cloud being 4~kpc. The observed $\gamma$-ray emission may also come from a combination of the first two scenarios. The last option would require the detection of a nearby SNR, which has not yet been detected, making it less probable than the two other options.\\

One additional component is also needed to model the region: a large extended source of angular size $\sigma~=~1^{\circ}43
\pm 0^{\circ}17_{\mbox{\scriptsize stat}}$. This may indicate either the mismodeling of 3HWC~J1928+178 or the lack of a large-scale galactic diffuse emission component in the model. We checked the expected flux of the galactic diffuse emission underlying the three sources J1928, J1930, and J1932 by using four different models: the latest Fermi model for Pass 8 and source class events~\citep{4FGL}, the diffuse emission model developed to simulate the Galactic plane survey with the future Cherenkov Telescope Array (CTA)~\citep{CTAdifuseEmission}, an updated version of this model (private communication with Q. Remy), and a model of the galactic diffuse emission up to 100 TeV developed by~\citet{PeVdiffuseEmission}.
On average, the contribution from J1928-EXT to these sources is more than twice the average flux that is expected from the galactic diffuse emission. 
This implies either that the diffuse emission does not represent the emission well, or that J1928-EXT contains more signal than diffuse emission, some of which may be left over from J1928, for example. If PSR~J1928+178 is responsible for this component, then the $\gamma$-ray luminosity would be $L_{\gamma} = 7.2 \times 10^{34}$~erg~s$^{-1}$. Assuming that these $\gamma$ rays are produced by IC scattering on CMB photons, the energy density would less than $\epsilon_{\mbox{\tiny IC}} = 0.001$~eV~cm$^{-3}$. Another hypothesis is that the two extended components J1928 and J1928-EXT may be the same object, if we consider a diffusion model similar to that of Geminga~\citep{3HWCJ1928_ICRC}. This hypothesis would favor the TeV halo nature of 3HWC~J1928+178.
Deeper analysis will be required to determine whether it can be related to existing sources, whether it comes from other sources, or whether it comes from large-scale $\gamma$-ray galactic emission.\\
 
Going farther will require better energy and angular resolutions: future analysis with energy estimators~\citep{HAWC_high_energy}, together with more data, would be appropriate for allowing a better study of the energy dependence of the spectral and morphological parameters. Moreover, a better angular resolution would permit the making profiles in different directions around the pulsar, along the cloud location and perpendicular to it, to see whether there is any asymmetry in the $\gamma$-ray emission.

%% IMPORTANT! The old "\acknowledgment" command has be depreciated. It was
%% not robust enough to handle our new dual anonymous review requirements and
%% thus been replaced with the acknowledgment environment. If you try to 
%% compile with \acknowledgment you will get an error print to the screen
%% and in the compiled pdf.

\section*{Acknowledgments}
\small {We acknowledge support from: the US National Science Foundation (NSF); the US Department of Energy Office of High-Energy Physics; the Laboratory Directed Research and Development (LDRD) program of Los Alamos National Laboratory; Consejo Nacional de Ciencia y Tecnolog\'ia (CONACyT), M\'exico, grants Nos. 271051, 232656, 260378, 179588, 254964, 258865, 243290, 132197, A1-S-46288, and A1-S-22784, c\'atedras 873, 1563, 341, and 323, Red HAWC, M\'exico; DGAPA-UNAM grants Nos. IG101320, IN111716-3, IN111419, IA102019, IN110621, and IN110521; VIEP-BUAP; PIFI 2012, 2013 and PROFOCIE 2014, 2015; the University of Wisconsin Alumni Research Foundation; the Institute of Geophysics, Planetary Physics, and Signatures at Los Alamos National Laboratory; Polish Science Centre grant No. DEC-2017/27/B/ST9/02272; Coordinaci\'on de la Investigaci\'on Cient\'ifica de la Universidad Michoacana; Royal Society - Newton Advanced Fellowship 180385; Generalitat Valenciana, grant No. CIDEGENT/2018/034; the Program Management Unit for Human Resources \& Institutional Development, Research and Innovation, NXPO (grant No. B16F630069); Coordinaci\'on General Acad\'emica e Innovaci\'on (CGAI-UdeG), PRODEP-SEP UDG-CA-499; and the Institute of Cosmic Ray Research (ICRR), University of Tokyo. H.F. acknowledges support from NASA under award No. 80GSFC21M0002. We also acknowledge the significant contributions over many years of Stefan Westerhoff, Gaurang Yodh, and Arnulfo Zepeda Dominguez, all deceased members of the HAWC collaboration. Thanks to Scott Delay, Luciano D\'iaz, and Eduardo Murrieta for technical support.}

%% To help institutions obtain information on the effectiveness of their 
%% telescopes the AAS Journals has created a group of keywords for telescope 
%% facilities.
%
%% Following the acknowledgments section, use the following syntax and the
%% \facility{} or \facilities{} macros to list the keywords of facilities used 
%% in the research for the paper.  Each keyword is check against the master 
%% list during copy editing.  Individual instruments can be provided in 
%% parentheses, after the keyword, but they are not verified.

\vspace{5mm}
\facility{HAWC (https://www.hawc-observatory.org/)}

%% Similar to \facility{}, there is the optional \software command to allow 
%% authors a place to specify which programs were used during the creation of 
%% the manuscript. Authors should list each code and include either a
%% citation or url to the code inside ()s when available.

\software{naima~\citep{naima}, 3ML~\citep{3ML}}

%% Appendix material should be preceded with a single \appendix command.
%% There should be a \section command for each appendix. Mark appendix
%% subsections with the same markup you use in the main body of the paper.

%% Each Appendix (indicated with \section) will be lettered A, B, C, etc.
%% The equation counter will reset when it encounters the \appendix
%% command and will number appendix equations (A1), (A2), etc. The
%% Figure and Table counter will not reset.

\newpage
\appendix
\section{Multiwavelength information}
\label{Multi-wavelength information}
Tables~\ref{J1930_caracteristics}, \ref{J1928_caracteristics} and \ref{J1932_caracteristics} gather the information regarding the three sources 3HWC J1930+188, 3HWC J1928+178 and HAWC J1932+192 in different wavelengths found in the literature, with the associated references.

%\vspace{1cm}

\begin{table}[ht!]
 \centering
 \begin{tabular}{|c|c c |c|c|c|} 
  Component            & \multicolumn{2}{c|}{Observations}                              &  Parameter          & Value   & Comments and references     \\ 
  \hline
%%%%%%%%%%%%%%%%%%%%%%%%%%%%%%%%%%%%%%%%%%%%%%%%%%%%%%%%%%%%%%%%%%%%%%%%%%%%%%%%%%%%%%%%%%%%%%%%%%%%%%%%%%%%%%%%%%%%%%%%%%%%%%%%%%%%%%%%%%%%%%%%%%%%%%%%%%%%
                       & \multirow{5}*{Radio}         & \multirow{5}*{Arecibo}          & period P (ms)       & 137            &  \multirow{5}*{\cite{Discovery_PSRJ1928}}  \\
                       &                              &                                 & $\dot{P}$    & $7.5\times10^{-13}$ &    \\
%%%%%%%%%%%%%%%%%%%%%%%%%%%%%%%%%%%%%%%%%%%%%%%%%
     Pulsar           &                              &                                 & $\dot{E}$ (erg~s$^{-1}$) & $12\times10^{36}$ & \\
                      &                              &                                 & age (kyr)           & 2.9           &     \\
   PSR J1930+1852       &                              &                                 & surf. B field (G) & $1.0\times10^{13}$ &     \\
\cline{2-6}           &  \multirow{5}*{X-ray}   & \multirow{5}*{\textit{Chandra}} & F (0.3-10 keV)    & $2.1\times10^{-12}$ & \\
                       &                              &                                 & index               & $-1.44 \pm 0.04 $ &  Pulsar, ring, jet, and  \\
\cline{1-1}\cline{4-5} &                              &                                 & size (\textdegree) & 0.03$\times$0.02 & diffuse elongated PWN \\
                       &                              &                                 & F (0.3-10 keV)     & $1.18\times10^{-12}$ & \cite{G54_Chandra_Spiter} \\
                       &                              &                                 & index              & $-2.2 \pm 0.04$ &   \\
%%%%%%%%%%%%%%%%%%%%%%%%%%%%%%%%%%%%%%%%%%%%%%%%%
  \cline{2-6}          & \multirow{4}*{Radio}         & \multirow{2}*{Effelsberg}    & \multirow{2}*{size (\textdegree)} & \multirow{2}*{0.025} & \multirow{2}*{\cite{G54_Effelsberg}} \\
                       &                              &                              &                                   &                    &          \\
  \cline{3-6}          &                              & \multirow{2}*{FCRAO}         & \multirow{2}*{distance (kpc)}     & \multirow{2}*{6.2} &  Association with a molecular cloud \\  
                       &                              &                              &                                   &                    & \cite{PWNG54_CO} \\
  \cline{2-6}G54.1+0.3 & \multirow{9}*{$\gamma$-ray} & \multirow{2}*{\textit{Fermi}}& \multirow{2}*{} & \multirow{2}*{} & Detection of a point-like source \\
                       &                              &                              &                                   &               & consistant with the VERITAS measurements  \\  
  \cline{3-5}          &                              & \multirow{2}*{VERITAS}       & index                             & $-2.18 \pm 0.2 $ &  \cite{Veritas_Fermi_2HWCsources} \\
                       &                              &                              & flux 1-100 TeV                    & $(3.31 \pm 1.47 )\times10^{-12}$  &        \\
  \cline{3-6}          &                              & \multirow{3}*{H.E.S.S.}      & size (\textdegree)                & 0.02 $\pm$ 0.025 & \multirow{3}*{\cite{HGPS}} \\ 
                       &                              &                              & index                             & $-2.59 \pm 0.26$ &         \\
                       &                              &                              & flux 1-100 TeV                    & $(1.28 \pm 0.55) \times10^{-12}$  &         \\       
  \cline{3-6}          &                              & \multirow{2}*{HAWC}          & index                             & $-2.76 \pm 0.14$  & \multirow{2}*{\cite{3HWC_catalog}}  \\ 
                       &                              &                              & flux 1-100 TeV                    & $(4.48 \pm 0.43) \times10^{-12}$ &             \\      
%%%%%%%%%%%%%%%%%%%%%%%%%%%%%%%%%%%%%%%%%%%%%%%%%%%%%%%%%%%%%%%%%%%%%%%%%%%%%%%%%%%%%%%%%%%%%%%%%%%%%%%%%%%%%%%%%%%%%%%%%%%%%%%%%%%%%%%%%%%%%%%%%%%%%%%%%%%%
\cline{1-6}            & radio                 & VLA                             & size (\textdegree)               & 0.1           &  \cite{G54_VLA}\\
\cline{2-6}            & Sub-mm                & \textit{Herschel}               & dust mass ($M_{\odot}$)          & 0.08 - 0.9    &  \multirow{3}*{\cite{G54_Hershel}}\\
 Shell                 & \multirow{3}*{IR}     & \multirow{3}*{\textit{Spitzer}} & dust temperature (K)             & 27 - 44       &  \\ 
                     &                       &                                 & progenitor's mass ($M_{\odot}$)  &  15 - 27      &   \\
   SNR G54.1+0.3     &                       &                                 & size (\textdegree)               & 0.4           &  \cite{G54_Spitzer} \\
                       
\cline{2-6}            &  \multirow{2}*{X-ray} &  \textit{XMM}                   & size (\textdegree)               &  $\sim$ 0.1 &  \multirow{2}*{\cite{G54_age}}  \\ 
                       &                       &  \textit{Suzaku}                & age (kyr)                        & 1.8 - 2.4   &   \\     
  \hline
  \hline   
\end{tabular}
\caption{ Characteristics of the components associated with 3HWC J1930+188 - Fluxes are in erg~cm$^{-2}$~s$^{-1}$}
\label{J1930_caracteristics}
\end{table}

\newpage

\begin{table}[ht!]
 \centering
 \begin{tabular}{|c|c c |c|c|c|} 
  Component            & \multicolumn{2}{c|}{Observations}                    &  Parameter          & Value   & Comments      \\ 
  \hline
%%%%%%%%%%%%%%%%%%%%%%%%%%%%%%%%%%%%%%%%%%%%%%%%%%%%%%%%%%%%%%%%%%%%%%%%%%%%%%%%%%%%%%%%%%%%%%%%%%%%%%%%%%%%%%%%%%%%%%%%%%%%%%%%%%%%%%%%%%%%%%%%%%%%%%%%%%%%
                       & \multirow{6}*{Radio} & \multirow{6}*{Arecibo}       & period P (ms)       & 68.7            & \multirow{6}*{\cite{Discovery_PSRJ1928}} \\
                       &                      &                              & $\dot{P}$    & $1.32\times10^{-14}$ &    \\
 Pulsar                &                      &                              & $\dot{E}$ (erg~s$^{-1}$)& $1.6\times10^{36}$ & \\
  PSR J1928+1746     &                      &                              & age (kyr)           & 82              &     \\
                     &                      &                              & distance (kpc)      & 4.3             &         \\
                      &                      &                              & surf. B field (G)   & $9.6\times10^{11}$  &     \\
\cline{1-6}  \multirow{4}*{PWN}   & \multirow{4}*{$\gamma$-ray}  & \multirow{2}*{EGRET} & index & $-2.23$ & \multirow{2}*{\cite{3Egret_catalog}} \\ 
                       &                      &                              & flux $>$ 100 MeV (ph~cm$^{-2}$~s$^{-1}$) & $157\times10^{-8}$ &       \\
\cline{3-6}            &                      & \multirow{2}*{HAWC}          & index                                    & $-2.3 \pm 0.07$ &  \multirow{2}*{\cite{3HWC_catalog}} \\   
                       &                      &                              & flux 1-100 TeV                           & $(4.77 \pm 0.32) \times 10^{-12}$ &     \\
  \hline
  \hline 
%%%%%%%%%%%%%%%%%%%%%%%%%%%%%%%%%%%%%%%%%%%%%%%%%%%%%%%%%%%%%%%%%%%%%%%%%%%%%%%%%%%%%%%%%%%%%%%%%%%%%%%%%%%%%%%%%%%%%%%%%%%%%%%%%%%%%%%%%%%%%%%%%%%%%%%%%%%%
\end{tabular}
\caption{ Characteristics of the components associated with 3HWC J1928+178 -  Fluxes are in erg~cm$^{-2}$~s$^{-1}$.}

\label{J1928_caracteristics}
\end{table}

\vspace{1cm}

\begin{table}[ht!]
 \centering
 \begin{tabular}{|c|c c |c|c|c|} 
  Component            & \multicolumn{2}{c|}{Observations}                              &  Parameter          & Value   & Comments      \\ 
  \hline
%%%%%%%%%%%%%%%%%%%%%%%%%%%%%%%%%%%%%%%%%%%%%%%%%%%%%%%%%%%%%%%%%%%%%%%%%%%%%%%%%%%%%%%%%%%%%%%%%%%%%%%%%%%%%%%%%%%%%%%%%%%%%%%%%%%%%%%%%%%%%%%%%%%%%%%%%%%%
                       & \multirow{8}*{$\gamma$-ray}  & \multirow{8}*{\textit{Fermi}}  & period P (ms)          &   208          & Radio-quiet \\
                       &                              &                       & max distance (kpc)              &  6.6 & Assuming 100\% efficiency in $\gamma$ rays \\
                       &                              &                       & $\dot{E}$ (erg~s$^{-1}$)        &  $4.07\times10^{35}$ & \\
                       &                              &                       & age (kyr)                       &  35.4            & \\
                       &                              &                       & flux $>$ 100 MeV                &  $7.8\times10^{-11}$  & \cite{Discovery_PSRJ1932} \\
                       &                              &                       & cut-off energy (GeV)            & 1.2             &  \\              
Pulsar                 &                              &                       & index                           & $-1.7\pm0.1$     &  \\
PSR J1932+1916         &                              &                       & surf. B field (G)               &  $4.5\times10^{12}$  &  \\
\cline{2-6}            &  \multirow{7}*{X-ray} & \multirow{4}*{\textit{Swift}} & distance (kpc)                 & 2-6 & Based on interstellar extinction \\
\cline{4-6}            &                              &                       & size (\textdegree)              & $<$ 0.008  &   \\
                       &                              &                       & flux (0.5-5 keV)                & $1.3\times10^{-13}$   &  \\
                       &                              &                       & index                         & $-1.4 \pm 1.0$ & From morphological \\                       
%%%%%%%%%%%%%%%%%%%%%%%%%%%%%%%%%%%%%%%%%%%%%%%%%
\cline{1-1}\cline{3-5} \multirow{3}*{PWN}    &                            & \multirow{3}*{\textit{Suzaku}} &  size (\textdegree)     & 0.075 & and spectral fit \\
                       &                              &                       & flux (0.5-5 keV)                & $1.2\times10^{-12}$   &   \cite{J1932_Xrays} \\
                      &                              &                       &  index                          & $-1.8 \pm 0.4$     &   \\
%%%%%%%%%%%%%%%%%%%%%%%%%%%%%%%%%%%%%%%%%%%%%%%%%
\cline{1-6}               & \multirow{5}*{Radio} & \multirow{2}*{Arecibo}        & distance (kpc)           & 3 - 4 & Association with CO cloud \\
                          &                      &                               & dynamical age (kyr)      & 95    & \cite{G54.4}   \\
\cline{3-6} Shell         &                      &  VLA                          & distance (kpc)           & 6.6   & Using H$_{\mbox{I}}$ absorption spectra \\
 SNR G54.4-0.3            &                      &  FCRAO                        & dynamical age (kyr)      & 190   & \cite{G54.4_HI}   \\
\cline{3-6}               &                      &                               & size (\textdegree)       & 0.67  & \cite{Green_catalog}  \\
  \hline
  \hline   
\end{tabular}
\caption{ Characteristics of the components associated with HAWC J1932+192 - Fluxes are in erg~cm$^{-2}$~s$^{-1}$}

\label{J1932_caracteristics}
\end{table}

%% For this sample we use BibTeX plus aasjournals.bst to generate the
%% the bibliography. The sample631.bib file was populated from ADS. To
%% get the citations to show in the compiled file do the following:
%%
%% pdflatex sample631.tex
%% bibtext sample631
%% pdflatex sample631.tex
%% pdflatex sample631.tex

\bibliographystyle{aasjournal}
\bibliography{biblio}{}

%% This command is needed to show the entire author+affiliation list when
%% the collaboration and author truncation commands are used.  It has to
%% go at the end of the manuscript.
%\allauthors

%% Include this line if you are using the \added, \replaced, \deleted
%% commands to see a summary list of all changes at the end of the article.
%\listofchanges

\end{document}